\documentclass[twocolumn,10pt]{article}

\usepackage[utf8]{inputenc}
\usepackage[T1]{fontenc}
\usepackage{amsmath,amsfonts,amssymb,amsthm}
\usepackage{graphicx}
\usepackage{xcolor}
\usepackage{tcolorbox}
\usepackage{hyperref}
\usepackage{cite}
\usepackage{authblk}
\usepackage{geometry}
\usepackage{booktabs}
\usepackage{multirow}
\usepackage{tabularx}
\usepackage{url}
\usepackage{algorithm}
\usepackage{algpseudocode}
\usepackage{tikz}
\usepackage{pdflscape}
\usetikzlibrary{shapes,arrows,positioning,calc,patterns,decorations.pathreplacing}
\usetikzlibrary{arrows.meta,fit,backgrounds}

\usepackage{colortbl}
\usepackage{pifont}
\usepackage{array}
\usepackage{ragged2e}
\usepackage{fancyvrb}
\usepackage{listings}
\usepackage{enumitem}
\usepackage{caption}
\usepackage{subcaption}
\usepackage{balance}

\definecolor{headerblue}{RGB}{37, 63, 114}      
\definecolor{tlsband}   {RGB}{232, 240, 253}    
\definecolor{ipsecband} {RGB}{232, 246, 236}    
\definecolor{sbiband}   {RGB}{253, 244, 230}    
\definecolor{monband}   {RGB}{245, 236, 250}    
\definecolor{fuzzband}  {RGB}{253, 233, 233}    
\definecolor{benchband} {RGB}{238, 238, 238}    

\definecolor{preqband}  {RGB}{238, 246, 255}    
\definecolor{hybl1band} {RGB}{229, 243, 229}    
\definecolor{hybl3band} {RGB}{200, 226, 197}    
\definecolor{hybl5band} {RGB}{163, 207, 161}    
\definecolor{pql1band}  {RGB}{247, 233, 255}    
\definecolor{pql3band}  {RGB}{226, 198, 247}    
\definecolor{pql5band}  {RGB}{195, 154, 229}    
\definecolor{verdictok} {RGB}{224, 242, 224}    
\definecolor{verdictwarn}{RGB}{255, 241, 210}   
\definecolor{verdictbad} {RGB}{255, 225, 225}   

\definecolor{tlsblue}{HTML}{1A4D8F}
\definecolor{lightgrayrow}{gray}{0.95}
\definecolor{pqpurple}{HTML}{6B21A8}
\definecolor{codegreen}{rgb}{0.0,0.4,0.0}
\definecolor{codegray}{rgb}{0.5,0.5,0.5}

\newcommand{\cmark}{\textcolor{tlsblue}{\ding{51}}}
\newcommand{\xmark}{\textcolor{red!70!black}{\ding{55}}}
\newcommand{\pqval}{\textsc{PQC Validator}}

\hypersetup{colorlinks=true,linkcolor=blue,citecolor=blue,urlcolor=cyan}

\newtheorem{theorem}{Theorem}

\newtheorem{definition}[theorem]{Definition}
\newtheorem{proposition}[theorem]{Proposition}

\title{\textbf{PQC Validator: Validating Post-Quantum Readiness in Cloud-Native 5G Core Networks}}
\author[1]{Lakshya Chopra} 
\author[2]{Vipin Kumar Rathi}

\affil[1]{coRAN Labs Private Limited, New Delhi, India}
\affil[2]{Ramanujan College, University of Delhi, New Delhi, India}

\date{}

\begin{document}

\twocolumn[
\begin{@twocolumnfalse}
\maketitle
\begin{abstract}
5G Core networks are entering a decisive phase of post-quantum (PQ) migration:
operators and vendors are beginning to advertise PQ-TLS 1.3, PQ-IPsec, and
hybrid KEM support across the Service-Based Interface (SBI) and N2/N3/N4
reference points, in line with 3GPP TS~33.501, emerging IETF drafts, and
NIST FIPS~203, 204, 205. Yet deploying PQ primitives does not guarantee PQ
security. A Network Function may advertise ML-KEM-768 and silently fall back
to X25519; negotiate a hybrid KEM but authenticate with ECDSA-P256; present
an ML-DSA leaf on a classical chain; or skip mutual TLS altogether. These
failures are silent on the wire, and today's scanners (testssl.sh, sslyze,
Qualys) together with 5G-specific fuzzers are PQ-unaware and telecom-blind.

We present \pqval{}, a layered PQC assurance framework purpose-built for the cloud-native 5G Core, comprising a PQ Crypto Engine (L1), a PQ Conformance Prober (L2), a PQ Robustness Tester (L3), a PQ Overhead Meter (L4), and an eBPF Attestation Plane for wire-level ground truth. Its scope spans the full
control-plane cryptographic surface: an independent PQ-TLS~1.3 client and
server, a strongSwan-driven PQ-IPsec harness for N2/N3/N4, an eBPF/XDP/TC
monitoring plane that extracts wire-level ground truth on negotiated groups
and signatures, and a Kubernetes-native UI that auto-discovers NFs and emits
structured \emph{PQ evidence} classifying every endpoint as
\texttt{classical}, \texttt{hybrid-pq}, or \texttt{full-pq}. A compliance
suite spans TLS, PQC, 3GPP SBI, NRF OpenAPI, and security hardening,
while a protocol fuzzer exercises CVE-class regressions and downgrade
paths. \pqval{} closes the gap between \emph{claimed} and
\emph{actual} PQ readiness, giving operators the independent, reproducible
evidence they need before trusting PQ deployments in production. We report
preliminary results against QORE, an open source Post Quantum 5G core, and outline a
roadmap toward formally verified KATs, a PQ primitive verifier, and PQ-PKI
conformance for 3GPP migration profiles.
\end{abstract}

\vspace{1.5em}
\vspace{1.5em}

\end{@twocolumnfalse}]

\section{Introduction}

The looming threat of quantum computing to classical cryptographic methods has persuaded organizations worldwide to carry out post-quantum migration of their cryptographic stacks. Coupled with the standardization and development of post-quantum methods (e.g., ML-KEM), the migration process has picked up pace, with recent reports citing that more than 40\%  of the internet traffic is now protected using \texttt{X25519MLKEM768} (a hybrid post-quantum KEM). Several cloud providers now offer Post-quantum TLS and Key management systems (KMS), which aim to mitigate the "Harvest Now, Decrypt later" attack. Open source libraries, such as OpenSSL, have added support for post-quantum primitives in normal cryptographic operations (e.g., key generation, signing). OpenSSL 3.5.0 provides Hybrid TLS 1.3 (which supports the usage of post-quantum hybrid KEMs). The migration procedure is also fueled by several internet drafts and RFCs, which introduce new protocol formats, modifications, mechanisms, and adjustments to make, to cleanly integrate post-quantum schemes into existing security infrastructure (e.g., TLS 1.3, IPsec, PKIX, etc). For instance, post-quantum methods introduce larger key exchange and ciphertext sizes, which existing protocols, including IPsec, could not accommodate; subsequently, as a preparation for the migration process, several key RFCs such as RFC 9232, and 9370 were introduced, that aim to fix packet fragmentation issues, and allow future integration of multiple key exchanges (e.g., DH3072 + ML-KEM-768), which paves the way for PQ-IPsec. 
In 5G networks for example, security infrastructure is massive and spread across multiple domains. This includes a wide coverage of IPsec, and DTLS protected communications, PKI managed trust, (mutual)TLS and OAuth 2.0 for internal authentication and authorization purposes, with additional network security boundaries, and firewalls. On the user side, 5G includes methods for authentication via shared symmetric keys, typically using 5G-AKA, with 128-bit keys (which could be susceptible to quantum attacks). 
Since, 5G is a critical infrastructure network, which carries signalling and user-plane traffic for billions of subscribers, with a rapidly growing population of IoT,
URLLC, satellite-based coverage and enterprise-slice endpoints, it is considerably important to safeguard it from quantum threats. Standard bodies, such as 3GPP, have begun carrying out feasibility studies for integrating quantum-safe methods to 5G security mechanisms, which are also expected to be a baseline for B5G/6G networks. Reports published by GSMA for example, detail protocol-level modifications and offer deep insights into upgrading existing cryptographic stack to a quantum-secure one. Open-source projects such as QORE, and Q-RAN, provide reference implementations of post-quantum methods and protocols, such as PQ-TLS 1.3, PQ-IPsec, PQ-OAuth 2.0, etc. However, the integration of "new" cryptographic primitives invite considerable uncertainty, since ways to adequately test implementations, check security of such schemes and protocols are not yet well-developed.

Deploying PQ algorithms does not equate to being PQ-secure. A 5G core may
advertise support for ML-KEM-768 in its TLS 1.3 \texttt{supported\_groups} extension yet still continue to prefer X25519 or fallback to other methods. A server
may negotiate a hybrid KEM for key exchange while signing with classical
ECDSA-P256 in \texttt{CertificateVerify}, leaving authentication vulnerable.
Certificate chains may be signed entirely with classical algorithms even
when the leaf uses ML-DSA. Mutual TLS may be absent between NFs despite
TS~33.501 mandating it. Certificate verification might be off, and secrets might not be handled cleanly. These failure modes are silent: the TCP connection
succeeds, the TLS handshake completes, HTTP/2 traffic flows, and the
operator sees nothing amiss. With existing tools, we can get decently good coverage for classical methods, however, several of them lack PQ support yet. Furthermore, the lack of such validators, tuned for telecom, is a concerning point, and doesn't resolve our problem.

Operators would prefer incorporating the fledging security mechanisms of PQC into their production network only after sufficient trust has been gained in their security, implementation correctness, and performance. Keeping this in mind, we introduce a \textbf{PQC Validator}, that aims to validate PQC implementations (particularly for the \textit{telco} domain), evaluating whether they are truly post-quantum or not, offering benchmarking and fuzzing test suites, with additional compliance tests.

\section{Motivation}\label{sec:motivation}

There is today no general-purpose way to test post-quantum deployments at
\emph{telco scale}. Operators moving to PQ-TLS or PQ-IPsec face three
concrete gaps that directly obstruct upgrades.

\subsection{No PQ-Aware Scanners for 5GC}

Classical TLS scanners - testssl.sh, sslyze, Qualys SSL Labs - operate on
the assumption that all \texttt{NamedGroup} IDs and all
\texttt{SignatureScheme} IDs are classical. When presented with a server
advertising \texttt{0x11EC} (X25519+ML-KEM-768) or \texttt{0x0B03}
(ML-DSA-87), they either log ``unknown group'' or abort. None of them
produce PQ evidence, and none of them distinguish hybrid PQ from pure PQ.
5G-specific protocol testing tools (5GReasoner, DoLTEst, RANsacked) focus
on the RAN and NAS layers - fuzzing radio and non-access-stratum signaling
- and are entirely PQ-unaware; they have no notion of TLS \texttt{key\_share}
parsing or IKEv2 \texttt{IKE\_INTERMEDIATE} exchanges. This tooling gap
blocks operator confidence and stalls upgrades: without a neutral way to
check a deployment, teams fall back to vendor-reported claims, which is
precisely what security auditors refuse to accept.



\subsection{No Path to Operator Trust}

Without independent evidence, operators are forced to trust vendor claims.
Worse, a validator that reuses the same TLS library as the target collapses
into a tautology: a bug present in both sides will not be detected. For
example, if both the target NF and the validator rely on OpenSSL+liboqs,
and liboqs has a bug in ML-KEM-768 encapsulation that produces wrong but
mutually consistent shared secrets, both sides will derive the same (wrong)
keys, the TLS Finished MAC will verify, and the validator will happily
report success. Cross-implementation independence is therefore a design
requirement, not a preference.

\pqval{} addresses these three gaps by (a) shipping an independent
PQ-TLS/IPsec stack implemented in Go against CIRCL, decoupled from the
OpenSSL+liboqs ecosystem used by most 5G cores; (b) embedding 3GPP+IETF+NIST
test cases in a single compliance suite that reports against specific clauses;
and (c) producing machine-verifiable JSON evidence that can be archived
and audited. In the longer term, \pqval{} aims to offer KATs and a
full 5G-core test suite that is PQ-compatible, giving operators a
reproducible basis for trust as they roll out PQ across their fleets.
Table~\ref{tab:failure-modes} summarizes the silent failure modes we
target; the complete test inventory is in
Appendix~\ref{app:test-sheet}.

\begin{table}[t]
\centering
\caption{Silent PQ failure modes that \pqval{} detects.}
\label{tab:failure-modes}
\small
\renewcommand{\arraystretch}{1.15}
\begin{tabularx}{\columnwidth}{@{}l X l@{}}
\toprule
\textbf{Dimension} & \textbf{Failure mode} & \textbf{Suite} \\
\midrule
\rowcolor{lightgrayrow}
PQ Capability  & KEX falls back to classical group      & PQC \\
PQ Capability  & Sig is ECDSA despite PQ KEX            & PQC \\
\rowcolor{lightgrayrow}
PQ PKI         & Cert chain uses classical sigs         & PQC \\
TLS Compliance & Weak cipher accepted (CBC, RC4)        & TLS \\
\rowcolor{lightgrayrow}
TLS Compliance & Expired or self-signed leaf            & TLS \\
5G Conformance & mTLS absent on SBI endpoint            & SBI \\
\rowcolor{lightgrayrow}
5G Conformance & HTTP/1.x downgrade permitted           & SBI \\
NRF            & \texttt{nfType} missing in profile     & NRF \\
\rowcolor{lightgrayrow}
IPsec          & IKE falls back to non-PQ group         & IPsec \\
IPsec          & ESP uses AES-CBC instead of GCM        & IPsec \\
\bottomrule
\end{tabularx}
\end{table}

\section{Related Work}\label{sec:related}

Our work touches on several core domains, including post-quantum protocol implementations, 5G security testing, eBPF-based network monitoring and observability, and formal verification of implementations.   \textbf{Post-Quantum} implementations of real-world protocols has been addressed in several works, such as the Open Quantum Safe (OQS) ~\cite{oqs-provider} and OpenSSL projects ~\cite{openssl}. AWS-LC ~\cite{aws-lc}, and BoringSSL ~\cite{boringssl} besides several major cryptographic libraries have also introduced quantum-safe primitives, thereby providing hybrid TLS 1.3 stacks, which primarily use \texttt{X25519MLKEM768}. Works such as \cite{sosnowski2023pqtls} and \cite{kampanakis2024ttlb} evaluate the performance of PQ-TLS 1.3 in several network scenarios (e.g., high bandwidth, high loss, LTE/5G, etc), which play an important role in understanding PQC's impact. On the other hand, strongswan ~\cite{strongswan} now supports Post-Quantum IPsec, using PPKs, and Multiple Key exchanges as described the RFC 8784 and RFC 9370.

\subsection{5G Security Testing}

RANsacked~\cite{bennett2024ransacked} introduces a fuzzing framework
that targets cellular interfaces exposed to the base station and the
UE, and applies it to seven open-source and commercial LTE/5G cores,
uncovering 119 vulnerabilities (93 CVEs) enabling denial-of-service
and memory-corruption attacks. They also present \texttt{ASNFuzzGen},
a structure-aware fuzzer generator for arbitrary ASN.1 specifications
used across cellular interfaces. Earlier work fuzzes core networks
through UE-accessible interfaces~\cite{johansson2014tfuzz}, or mutates
messages exchanged between a legitimate UE and base
station~\cite{garbelini2022ota,potnuru2021berserker}. Chlosta et
al.~\cite{chlosta2021lteautomata} study the challenges of
reconstructing the NAS protocol state machine via automata learning
and use the resulting models to test several UE--MME implementations.


\paragraph{TLS testing tools.} testssl.sh, sslyze, and Qualys SSL Labs are
widely used for classical TLS posture assessment. Their cipher databases,
NamedGroup tables, and reporting rubrics have no PQ entries; PQ sessions
either show up as ``unknown'' or trigger parse errors. None of them are
Kubernetes-native and none perform NF-aware service discovery.

\subsection{Formal verification of TLS}
Project Everest~\cite{bhargavan2017everest} builds high-assurance
communication stacks on top of F*~\cite{swamy2016fstar}, a
proof-oriented dependently typed language in which programs are
written alongside machine-checked proofs of their correctness and
security. Out of this effort come miTLS, a verified reference
implementation of TLS~1.3, together with the verified cryptographic
libraries HACL*~\cite{zinzindohoue2017haclstar} and
EverCrypt~\cite{protzenko2020evercrypt}, components of which are now
deployed in production systems such as Mozilla Firefox, the Linux
kernel, and WireGuard. More recently,
Bertie~\cite{bhargavan2025bertie} provides a post-quantum TLS~1.3
implementation in Rust whose security and functional correctness are
established by translating the source via the \texttt{hax} toolchain
into F* and ProVerif, relying on libcrux for verified PQ primitives.
These efforts target classical or narrowly scoped PQ properties in
isolation, rather than 5G-integrated stacks or runtime deployment
validation; our roadmap connects \pqval{} to these verified libraries
so the validator can serve as a reference probe against a formally
trusted baseline.

\subsection{eBPF for Network Observability}
Several eBPF-based projects provide in-kernel visibility and policy
enforcement for cloud-native workloads. Cilium~\cite{cilium} delivers
networking, load balancing, and policy for Kubernetes, with
Hubble~\cite{hubble} exposing L3--L7 flow visibility, service maps,
policy observability, and exportable metrics. Tetragon~\cite{tetragon}
provides a security-focused stack that detects malicious patterns and
suspicious runtime behavior, and can enforce policies in-kernel.
Pixie~\cite{pixie} auto-instruments applications to expose
higher-layer signals such as HTTP and database queries in a
developer-friendly interface. KubeArmor~\cite{kubearmor} enforces
runtime security policies that constrain container behavior at the
system-call level. These tools offer broad visibility and enforcement,
but none classify cryptographic negotiation on the wire; \pqval{}'s
eBPF plane is narrower by design - it targets TLS and IKEv2
handshake parsing on the 5G reference points, producing PQ-specific
evidence rather than general-purpose flow metrics. We highlight that our validator's eBPF stack largely focuses on the impact introduced by post-quantum algorithms on the network-level, such as TCP congestions, multiple segments for client hello, increased fragmentations, multiple cert chain records, and more.

\begin{table}[t]
\caption{Comparison of TLS scanners against a PQ-TLS endpoint
(QORE AMF, X25519+ML-KEM-768 + ML-DSA-65). \cmark{} supported,
\xmark{} not supported.}
\label{tab:scanner-comparison}
\footnotesize
\setlength{\tabcolsep}{3pt} 
\renewcommand{\arraystretch}{1.05}

\begin{tabularx}{\columnwidth}{@{}p{0.48\columnwidth} *{4}{>{\centering\arraybackslash}X}@{}}
\toprule
\textbf{Capability} & 
\rotatebox{70}{testssl.sh} &
\rotatebox{70}{sslyze} &
\rotatebox{70}{Qualys} &
\rotatebox{70}{\pqval{}} \\
\midrule
\rowcolor{lightgrayrow}
Detect PQ NamedGroup IDs    & \xmark & \xmark & \xmark & \cmark \\
Detect PQ SignatureScheme   & \xmark & \xmark & \xmark & \cmark \\
\rowcolor{lightgrayrow}
Validate hybrid KEM         & \xmark & \xmark & \xmark & \cmark \\
PQ Evidence artifacts       & \xmark & \xmark & \xmark & \cmark \\
\rowcolor{lightgrayrow}
5G SBI test cases           & \xmark & \xmark & \xmark & \cmark \\
NRF OpenAPI validation      & \xmark & \xmark & \xmark & \cmark \\
\rowcolor{lightgrayrow}
K8s NF auto-discovery       & \xmark & \xmark & \xmark & \cmark \\
IPsec / IKEv2 tests         & \xmark & \xmark & \xmark & \cmark \\
\rowcolor{lightgrayrow}
eBPF wire attestation       & \xmark & \xmark & \xmark & \cmark \\
Own PQ-TLS stack            & \xmark & \xmark & \xmark & \cmark \\
\bottomrule
\end{tabularx}
\end{table}
\section{Background}\label{sec:background}

\subsection{Validators, Fuzzers, and Scanners}

We distinguish three kinds of active probes. A \emph{scanner} enumerates
exposed algorithms and configurations; its output is a list. A \emph{fuzzer}
injects malformed or mutated inputs to uncover crashes and memory safety
bugs; its output is a trace or signal. A \emph{validator} is a black-box
probe that actively negotiates a protocol and checks observable behavior
against a set of machine-enforceable rules; its output is structured
evidence that can be audited, archived, and compared across runs.
\pqval{} is primarily a validator, but it embeds a fuzzer and a scanner-like
benchmark module so that a single run produces conformance, robustness, and
performance artifacts in one pass.

\subsection{Post-Quantum Cryptography}

TLS~1.3 (RFC~8446) establishes a secure channel via the ClientHello /
ServerHello handshake followed by HKDF-based key derivation. Post-quantum
cryptography modifies two components of this handshake: \emph{key exchange}
- where a Key Encapsulation Mechanism (KEM) replaces (EC)DHE - and
\emph{authentication} - where PQ signatures replace RSA/ECDSA in
\texttt{CertificateVerify}.

\paragraph{ML-KEM (FIPS~203).} A Module-LWE-based KEM with three parameter
sets: ML-KEM-512, ML-KEM-768, and ML-KEM-1024 at NIST levels 1, 3, and 5.
Public keys are 800, 1184, and 1568 bytes; ciphertexts are 768, 1088, and
1568 bytes. Encapsulate/decapsulate operations replace Diffie-Hellman key
agreement.

\paragraph{ML-DSA (FIPS~204).} A Module-LWE-based signature scheme with
ML-DSA-44, ML-DSA-65, and ML-DSA-87 at levels 2, 3, and 5. ML-DSA-65
produces 3293-byte signatures, roughly 45$\times$ larger than ECDSA-P256.
Public keys are 1952 bytes for ML-DSA-65.

\paragraph{SLH-DSA (FIPS~205).} A stateless hash-based backup signature
with small keys but very large signatures (7--50~KB depending on parameter
set). Useful as a belt-and-suspenders option for long-lived root
certificates.

\paragraph{Hybrid groups.} During the transition, hybrid key exchange such
as X25519+ML-KEM-768 (draft-ietf-tls-ecdhe-mlkem, IANA \texttt{0x11EC})
combines a classical ECDH with a PQ KEM. The shared secret is derived from
the concatenation of both component secrets, so an attacker must break both
components to recover the key.

\paragraph{Wire-level impact.} PQ objects are large: an ML-KEM-768
public key is 1184~B, an ML-DSA-65 signature is 3293~B. A
PQ-TLS~1.3 ClientHello grows from $\approx$300~B to
$\approx$1400~B, and a hybrid key share is 1216~B instead of 32~B.
With Linux's default \texttt{TCP\_INIT\_CWND} of 10 segments and an
MSS of $\sim$1448~B on typical CNI veths, a classical ClientHello
fits in a single segment but a PQ one does not. For TLS over
low-MTU links (urban 4G/5G backhaul at MTU~1400, satellite at
1280), the handshake crosses TCP segmentation boundaries and the
cost is dominated by bytes-on-wire plus an extra RTT in case of
HelloRetryRequest, rather than by CPU.
Section~\ref{sec:ebpf} measures these effects directly via a
\texttt{sock\_ops} probe on \texttt{snd\_cwnd} and
\texttt{snd\_mss}.

\subsection{TLS 1.3 Handshake}

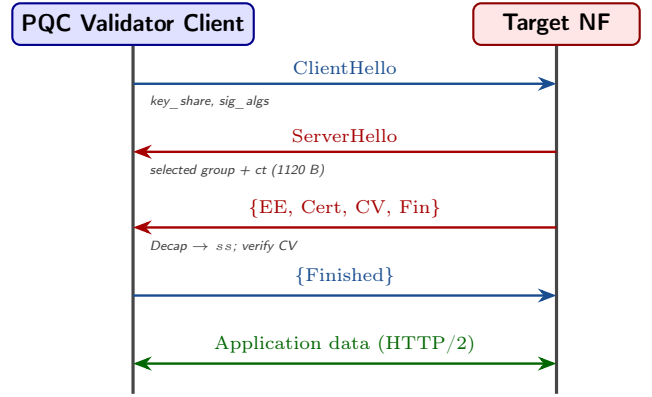
\begin{figure}[t]
\centering
\begin{tikzpicture}[
    >=Stealth,
    font=\scriptsize\sffamily,
    side/.style={
        draw, rounded corners=3pt, thick,
        minimum width=2.2cm, minimum height=0.55cm,
        text centered, font=\small\sffamily\bfseries
    },
    lifeline/.style={very thick, gray!55!black},
    msg/.style={->, thick, line width=0.9pt},
    appmsg/.style={<->, thick, line width=0.9pt, codegreen},
    opnote/.style={
        font=\tiny\sffamily\itshape, text=gray!50!black,
        align=left
    },
]

\node[side, fill=blue!12, draw=blue!55!black] (c) at (0,0)   {\pqval{} Client};
\node[side, fill=red!10,  draw=red!55!black]  (s) at (5.6,0) {Target NF};

\draw[lifeline] (c.south) -- ++(0,-4.6);
\draw[lifeline] (s.south) -- ++(0,-4.6);

\coordinate (m1c) at ($(c.south)+(0,-0.5)$);
\coordinate (m1s) at ($(s.south)+(0,-0.5)$);
\coordinate (m2c) at ($(c.south)+(0,-1.4)$);
\coordinate (m2s) at ($(s.south)+(0,-1.4)$);
\coordinate (m3c) at ($(c.south)+(0,-2.4)$);
\coordinate (m3s) at ($(s.south)+(0,-2.4)$);
\coordinate (m4c) at ($(c.south)+(0,-3.3)$);
\coordinate (m4s) at ($(s.south)+(0,-3.3)$);
\coordinate (m5c) at ($(c.south)+(0,-4.2)$);
\coordinate (m5s) at ($(s.south)+(0,-4.2)$);

\draw[msg, tlsblue] (m1c) -- 
    node[above, font=\scriptsize] {ClientHello} (m1s);
\node[opnote, anchor=west] at ($(m1c)+(0.1,-0.25)$)
    {key\_share, sig\_algs};

\draw[msg, red!65!black, <-] (m2c) -- 
    node[above, font=\scriptsize] {ServerHello} (m2s);
\node[opnote, anchor=west] at ($(m2c)+(0.1,-0.25)$)
    {selected group + ct (1120 B)};

\draw[msg, red!65!black, <-] (m3c) -- 
    node[above, font=\scriptsize] {\{EE, Cert, CV, Fin\}} (m3s);
\node[opnote, anchor=west] at ($(m3c)+(0.1,-0.25)$)
    {Decap $\to ss$;\ verify CV};

\draw[msg, tlsblue] (m4c) -- 
    node[above, font=\scriptsize] {\{Finished\}} (m4s);

\draw[appmsg] (m5c) -- 
    node[above, font=\scriptsize] {Application data (HTTP/2)} (m5s);

\end{tikzpicture}
\caption{PQ-TLS~1.3 handshake driven by \pqval{}. The client offers 
hybrid and pure-PQ \texttt{key\_share} groups together with PQ 
signature schemes, decapsulates the server's ciphertext, verifies 
\texttt{CertificateVerify} over the handshake transcript, and emits 
its own \texttt{Finished} before the SBI session begins. Per-stage 
timings, negotiated identifiers, and certificate-chain metadata are 
captured into the run's \texttt{PQEvidence}.}
\label{fig:handshake}
\end{figure}

Figure~\ref{fig:handshake} sketches the PQ-TLS~1.3 handshake that \pqval{}
drives. The ClientHello carries PQ groups in \texttt{supported\_groups},
PQ signatures in \texttt{signature\_algorithms}, a hybrid public key in
\texttt{key\_share}, and ALPN \texttt{h2} per TS~29.500. The server's
ServerHello selects a group and encapsulates against the client's public
key; ciphertext and signature are returned in the EncryptedExtensions and
CertificateVerify flights; a Finished MAC over the transcript closes the
handshake.

\subsection{5G Core Security Architecture}

\begin{figure}[t]
\centering
\begin{tikzpicture}[
    nf/.style={draw, rounded corners=2pt, minimum width=0.75cm, minimum height=0.45cm, font=\tiny\sffamily, text centered, fill=blue!10, draw=blue!50},
    rp/.style={draw, rounded corners=2pt, minimum width=0.9cm, minimum height=0.45cm, font=\tiny\sffamily, text centered, fill=red!10, draw=red!60},
    lbl/.style={font=\tiny, text=gray},
]
\node[nf] (nrf) at (3.2,1.6) {NRF};
\node[nf] (ausf) at (0,0.8) {AUSF};
\node[nf] (udm)  at (1.1,0.8) {UDM};
\node[nf] (pcf)  at (2.2,0.8) {PCF};
\node[nf] (nssf) at (3.3,0.8) {NSSF};
\node[nf] (nef)  at (4.4,0.8) {NEF};
\node[nf] (amf)  at (0.8,-0.2) {AMF};
\node[nf] (smf)  at (2.4,-0.2) {SMF};
\node[rp] (gnb)  at (0.8,-1.2) {gNB};
\node[rp] (upf)  at (3.5,-1.2) {UPF};
\node[rp] (dn)   at (5.2,-1.2) {DN};

\draw[gray!60] (nrf) -- (ausf); \draw[gray!60] (nrf) -- (udm);
\draw[gray!60] (nrf) -- (pcf);  \draw[gray!60] (nrf) -- (nssf);
\draw[gray!60] (nrf) -- (nef);  \draw[gray!60] (nrf) -- (amf);
\draw[gray!60] (nrf) -- (smf);
\draw[tlsblue, thick] (gnb) -- node[lbl,right]{N2/N3} (amf);
\draw[tlsblue, thick] (gnb) -- (upf);
\draw[pqpurple, thick] (smf) -- node[lbl,above]{N4} (upf);
\draw[codegreen, thick] (upf) -- node[lbl,above]{N6} (dn);

\node[lbl, anchor=west] at (4.8,1.6) {SBI};
\end{tikzpicture}
\caption{5G Core reference points. SBI (HTTP/2 over TLS 1.3) on the top
plane; N2/N3 (gNB, IPsec), N4 (PFCP, IPsec), and N6 (external) on the
bottom plane.}
\label{fig:5gc}
\end{figure}
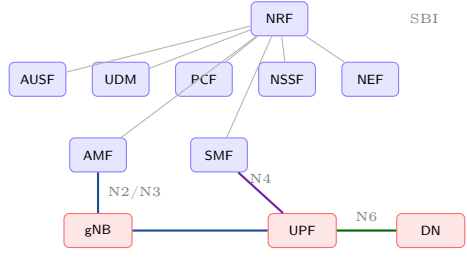

The 5G SBA comprises Network Functions communicating via HTTP/2-based SBI
(Figure~\ref{fig:5gc}). Reference points carry distinct traffic:

\begin{itemize}[leftmargin=*,itemsep=1pt]
    \item \textbf{SBI} (NF$\leftrightarrow$NF): HTTP/2 over TLS~1.3 with mTLS,
    IANA-registered ports 29500--29600.
    \item \textbf{N2} (gNB$\leftrightarrow$AMF): NGAP over SCTP (proto 132),
    protected by IPsec.
    \item \textbf{N3} (gNB$\leftrightarrow$UPF): GTP-U over UDP/2152,
    protected by IPsec.
    \item \textbf{N4} (SMF$\leftrightarrow$UPF): PFCP over UDP/8805,
    protected by IPsec.
\end{itemize}

3GPP TS~33.501 mandates TLS~1.3 with mTLS between NFs (\S6.3.4) and IPsec on
N2/N3/N4 (\S9.1) \cite{3gpp-ts-33501}. TS~29.500 \cite{3gpp-ts-29500} mandates HTTP/2 with ALPN \texttt{h2} and
prohibits HTTP/1.x fallback (\S6.1.3B). TS~29.510 defines the NRF service
discovery API; NF profiles must carry valid \texttt{nfType},
\texttt{nfStatus}, and PLMN identifiers. Critically, the specifications
mandate \emph{what} must hold, not \emph{how} operators should verify these
properties at runtime - the gap \pqval{} targets.

\subsection{QORE: A Post-Quantum 5G Core}

In this paper, we evaluate our validator by testing it against \textbf{QORE}, a cloud-native post-quantum 5G Core built over Aether-SDCore and free5GC. QORE is intended to serve as a reference architecture for post-quantum telecom networks, enabling operators to assess feasibility, conduct preliminary analysis, and perform hands-on testing with real user equipments and radios. QORE achieves quantum safety by integrating post-quantum primitives into golang's cryptographic stack, and additionally provides an external PQ-PKI for certificate issuance and trust management. We choose QORE as our primary validation target because it exposes post-quantum protocols and primitives, follows a cloud-native architecture, is open-source, and serves as a solid, trimmed-down counterpart to commercially-deployed core solutions. Our results are presented in Section~\ref{sec:results}.

\section{PQC Validator: Overview}\label{sec:overview}

In this section, we provide an overview of our validation framework, building it up from the cryptographic layer to the validation layer and beyond. We highlight our design principles and coverage scenarios. We note that our validator's primary focus remains the evaluation of post-quantum readiness in 5G core networks; however, as an extension of this work, we provide a complete 5G core security testing suite that includes several compliance test suites, offers detailed benchmarks, and ships with a (currently primitive) TLS fuzzer. Our presentation therefore considers the validator as a complete system, which we later break down into layers and individual tools.

Overall, our validator comprises the following, largely-independent subsystems:

\begin{enumerate}[leftmargin=*,itemsep=1pt]
    \item An \textbf{independent PQ-TLS 1.3 stack}, implemented in Golang, supporting several pure and hybrid post-quantum schemes (e.g., X25519MLKEM768 for key exchange, ML-DSA-65 for authentication). The stack supports mutual TLS when required, and exposes detailed insights from the handshake procedure - such as negotiated groups, derived secrets, handshake timings, and per-message sizes - which downstream modules consume for benchmarking and conformance checks.

    \item A \textbf{PQ-IPsec implementation}, based on RFC 8784 and RFC 9370, that allows users to test PQ-IPsec endpoints - typically running at the AMF (N2) or UPF (N3/N9) - using both post-quantum pre-shared keys and ML-KEM-768-based multiple key exchanges. Our custom implementation surfaces detailed statistics on protocol-level fragmentation, packet lengths, overhead relative to classical IKEv2, rekeying behavior, and negotiation failures.

    \item A \textbf{packet-monitoring subsystem} built using eBPF programs attached to tc/XDP hook points, which sniffs TLS and IKEv2 traffic in-kernel for packet classification, datapath analysis, bandwidth accounting, and determining whether a given flow is post-quantum or classical. This enables live observation of PQ traffic on a running core without modifying the network functions.

    \item A \textbf{multi-group compliance suite} (TLS, 5G~SBI, post-quantum security, NRF~OpenAPI, and security hardening) that checks whether the core network conforms to the security standards set by 3GPP~\cite{3gpp-ts-33501} and NIST~\cite{NIST-VPN-recs}. The suite covers cipher-suite negotiation, certificate validation, key-exchange group enforcement, session-resumption behaviour, and SBI-specific requirements such as OAuth2 token handling, mTLS enforcement, and HTTP/2 over TLS between NFs. A built-in TLS and IPsec fuzzer then stresses the implementation against malformed or adversarial inputs, and its findings are merged back into the hardening group. The full test sheet is reproduced in Appendix~\ref{app:test-sheet}.

    \item A \textbf{benchmarking and reporting module} that aggregates measurements from the above subsystems - handshake latency, CPU and memory footprint, bandwidth overhead, and message sizes - and emits structured, per-run JSON artifacts for downstream analysis and comparison across cores, versions, or cryptographic configurations.

    \item A \textbf{highly-configurable, Helm-based Kubernetes deployment} that packages the validator alongside the target 5G core, together with a web-based UI for launching test runs, inspecting live logs, and downloading artifacts, making the validation workflow accessible to operators without requiring deep familiarity with the individual tools.

    \item An \textbf{indicative comparison between classical and post-quantum protocols} exercised end-to-end against open-source cores. For each (KEX, Sig) combination the validator records handshake latency, per-stage cryptographic costs (KeyGen, Encap/Decap, Sign/Verify), and bytes-on-wire, and contrasts them against a classical baseline to surface the PQ overhead on a per-NF basis. A full datapath analysis --- involving many simultaneous UE attachments, a complete RAN$\leftrightarrow$Core path secured by mixed classical and post-quantum protocols, and sustained high-throughput traffic --- is out of scope for this paper and is left as future work; interpreting the results of~\cite{kampanakis2024ttlb,sosnowski2023pqtls}, we expect the per-handshake PQ cost to amortize over long-lived, HTTP/2-multiplexed SBI sessions and to play a negligible role in steady-state data-plane throughput.
\end{enumerate}

\paragraph{Design principles.} Four principles guide the design:
\emph{Cross-implementation interoperability} - the validator is implementation-agnostic: it works against any RFC-compliant TLS 1.3 or IKEv2 endpoint, and does not depend on internals of a specific 5G core distribution.
\emph{Evidence over reports} - every run produces downloadable, JSON-based artifacts alongside human-readable summaries, to ensure archivability, future auditing, reproducibility, and meaningful comparisons across runs.
\emph{5G-nativity} - discovery, classification, and reporting are specialized for the 3GPP service-based architecture (SBA) and its reference points, rather than retrofitted from a general-purpose web scanner.
\emph{Modularity} - each subsystem is usable in isolation (e.g., the PQ-TLS stack as a standalone client, or the eBPF monitor on an unmodified core), so operators can adopt the validator incrementally rather than all-at-once.

Table~\ref{tab:scope} summarizes scope.

\begin{table*}[t]
\centering
\caption{Scope of \pqval{}.}
\label{tab:scope}
\small
\renewcommand{\arraystretch}{1.2}
\begin{tabular}{lp{10cm}}
\toprule
\textbf{Dimension} & \textbf{Coverage} \\
\midrule
\rowcolor{lightgrayrow}
Protocols       & TLS 1.3 (RFC 8446); IKEv2/IPsec (RFC 7296, 8784, 9370); HTTP/2 over TLS (SBI) \\
KEMs            & \texttt{ML-KEM-512/768/1024; HQC-128/192/256}; hybrids: \texttt{X25519MLKEM768, P256MLKEM768, P384MLKEM1024} \\
\rowcolor{lightgrayrow}
Signatures      & \texttt{ML-DSA-44/65/87; SLH-DSA (SHA2/SHAKE, 128s/128f/192s/256s); Ed448+ML-DSA-65}; composite \texttt{ML-DSA+ECDSA}; classical baselines (\texttt{RSA, ECDSA, Ed25519}) \\
Authentication  & Server-only TLS; mutual TLS; IKEv2 PSK (RFC 8784); certificate-based IKEv2 \\
\rowcolor{lightgrayrow}
5G Interfaces   & SBI (HTTP/2 over TLS); N2 (SCTP/NGAP); N3 (GTP-U); N4 (PFCP); N6/N9 (observed via eBPF) \\
3GPP Standards  & TS 33.501 (security architecture); TS 33.210 (NDS/IP); TS 29.500, 29.510 (SBI); TS 23.003 \\
\rowcolor{lightgrayrow}
NIST / IETF     & FIPS 203/204/205; RFC 8446, 9370; draft-ietf-tls-ecdhe-mlkem; NIST SP 800-77r1 (IPsec VPN) \\
Test Inventory  & 50+ compliance tests across 5 groups; 16 fuzz cases; latency, throughput, and size benchmarks \\
\rowcolor{lightgrayrow}
Evidence        & Per-session PQ level; negotiated KEX/signature; certificate-chain PQ flag; handshake timings; JSON export \\
Deployment      & Host binary; Docker image; Helm chart (K8s-native, NodePort UI); CLI and web UI \\
\rowcolor{lightgrayrow}
Monitoring      & eBPF on TC/XDP; per-interface flow classifier; ring-buffer streaming to userspace collector \\
\bottomrule
\end{tabular}
\end{table*}

\subsection{Classifying PQ Security of a Session}
\label{subsec:pq-level}

For each observed TLS~1.3 session, \pqval{} records three pieces of evidence:
the negotiated key-exchange group, the signature scheme used in
\texttt{CertificateVerify}, and the certificate chain presented by the peer.
From these, it assigns one of three PQ-security labels per
Definition~\ref{def:pq-level}.

\begin{definition}[PQ security level of a TLS session]\label{def:pq-level}
Let $\mathcal{G}_\mathit{PQ}$ denote the set of post-quantum or
PQ-hybrid \texttt{NamedGroup} identifiers (e.g.\ \texttt{0x11EC},
\texttt{0x0768}) and $\mathcal{S}_\mathit{PQ}$ the set of post-quantum
or composite \texttt{SignatureScheme} identifiers (e.g.\ ML-DSA-65,
Ed448+ML-DSA-65). For a session that negotiates group $g$ and signs
\texttt{CertificateVerify} with scheme $\sigma$, we define the PQ
security level
$\ell \in \{\texttt{full-pq}, \texttt{hybrid-pq}, \texttt{classical}\}$
as:
\[
\ell \;=\;
\begin{cases}
\texttt{full-pq}   & \text{if } g \in \mathcal{G}_\mathit{PQ} \text{ and } \sigma \in \mathcal{S}_\mathit{PQ},\\[2pt]
\texttt{hybrid-pq} & \text{if } g \in \mathcal{G}_\mathit{PQ} \text{ and } \sigma \notin \mathcal{S}_\mathit{PQ},\\[2pt]
\texttt{classical} & \text{otherwise.}
\end{cases}
\]
The chain-level PQ status is reported as supplementary evidence and
does not alter $\ell$.
\end{definition}

The three labels are:

\begin{itemize}[leftmargin=*,itemsep=1pt]
    \item \textbf{\texttt{full-pq}} - both the key exchange and the
    server's signature use post-quantum (or PQ-hybrid) algorithms.
    \item \textbf{\texttt{hybrid-pq}} - the key exchange is post-quantum or
    hybrid, but the signature (typically tied to the certificate) is still
    classical. This is the dominant transitional state today, since PQ KEMs
    are deployable ahead of PQ PKI.
    \item \textbf{\texttt{classical}} - neither the key exchange nor the
    signature uses a post-quantum algorithm.
\end{itemize}

\noindent
Concretely, classification proceeds as a small lookup:
\begin{verbatim}
classify(session):
    g, sig, chain = session.kex_group,
                    session.cert_verify_sig,
                    session.cert_chain
    pq_kex = g   in PQ_NAMED_GROUPS
    pq_sig = sig in PQ_SIGNATURE_SCHEMES
    if   pq_kex and pq_sig: return "full-pq"
    elif pq_kex:            return "hybrid-pq"
    else:                   return "classical"
\end{verbatim}

The chain-level PQ status - whether any intermediate or root certificate
uses a PQ signature - is reported as supplementary evidence rather than
folded into the session label, since a PQ-signed leaf over a classical
root still meaningfully changes the quantum-attack surface and deserves
its own line in the report.

\begin{table*}[t]
\centering
\caption{Test suites grouped by subsystem. The validator ships
compliance tests across four protocol subsystems (PQ-TLS, PQ-IPsec,
SBI, and eBPF monitoring), a protocol fuzzer, and a configurable
benchmark harness. The full per-test catalogue is in
Appendix~\ref{app:test-sheet}.}
\label{tab:test-suites}
\renewcommand{\arraystretch}{1.5}
\setlength{\tabcolsep}{8pt}
\small
\begin{tabularx}{\textwidth}{@{}>{\bfseries}l c X@{}}
\toprule
\rowcolor{headerblue}
\textcolor{white}{\textbf{Subsystem}} &
\textcolor{white}{\textbf{Category}} &
\textcolor{white}{\textbf{Representative checks}} \\
\midrule

\rowcolor{tlsband}
PQ-TLS 1.3 & Compliance &
Negotiation of ML-KEM and hybrid groups (X25519MLKEM768, P256MLKEM768);
rejection of weak/legacy groups and cipher suites;
ML-DSA and composite ML-DSA+ECDSA signature verification;
session resumption with PQ parameters;
mTLS with PQ client certificates;
HelloRetryRequest handling for oversized key shares;
downgrade resistance (e.g., TLS 1.2 fallback attempts). \\

\rowcolor{ipsecband}
PQ-IPsec / IKEv2 & Compliance &
RFC 8784 post-quantum pre-shared key (PPK) negotiation;
RFC 9370 multiple key exchanges via \texttt{IKE\_INTERMEDIATE};
ML-KEM-768 additional key exchange;
fragmentation of oversized \texttt{IKE\_SA\_INIT} payloads;
rekeying behaviour under PQ-hybrid SAs;
combined PSK + PQ-KEM authentication;
PQ-aware DPD and SA teardown. \\

\rowcolor{sbiband}
SBI (HTTP/2 + TLS) & Compliance &
NF-to-NF mutual TLS enforcement;
OAuth2 access-token validation at the NRF / producer;
TLS profile conformance per TS 33.210;
\texttt{ALPN = h2} negotiation;
rejection of HTTP/1.1 and cleartext HTTP/2 (h2c) fallback;
SNI-based NF identity binding;
certificate-chain conformance to the 3GPP operator profile. \\

\rowcolor{monband}
Monitoring (eBPF) & Compliance &
Per-interface flow classification (TLS vs IKEv2 vs other);
PQ-vs-classical labelling of live handshakes;
TC and XDP program attachment on N2/N3/N6;
ring-buffer streaming without packet loss under load;
ClientHello and IKE\_SA\_INIT parsing at line rate;
bandwidth accounting per flow and per NF. \\

\rowcolor{fuzzband}
Fuzzing & Adversarial &
Malformed \texttt{ClientHello} / \texttt{ServerHello};
truncated and oversized key shares;
invalid / tampered ML-KEM ciphertexts;
IKEv2 payload length mismatches;
oversized SA proposals;
unexpected state transitions;
replayed and out-of-order handshake messages. \\

\rowcolor{benchband}
Benchmarks & Performance &
Handshake latency (p50 / p95 / p99);
CPU and memory cost during KEM encapsulation / decapsulation;
bytes-on-wire per handshake (ClientHello, ServerHello, Certificate);
sustained throughput under concurrent PQ sessions;
overhead relative to classical baseline (X25519, ECDSA P-256). \\

\bottomrule
\end{tabularx}
\end{table*}

\begin{proposition}[Soundness bounds]
If \pqval{} reports that a session negotiated $g$ and $\sigma$, the server
did select them in its ServerHello and CertificateVerify for this session:
soundness follows from performing a real handshake and Finished-MAC
verification. \pqval{} does \emph{not} verify policy: \texttt{full-pq}
means the server \emph{can}, not \emph{always will}, negotiate full PQ.
\end{proposition}

\section{Layered Architecture}\label{sec:arch}

\pqval{} is organised as a layered architecture, where each layer
consumes only the APIs and interfaces exposed by the layers below
it. For example, the \textbf{PQ Robustness Tester} (L3) uses the
TLS/IPsec connection methods exposed by the \textbf{PQ Conformance
Layer} (L2); the Conformance Layer itself stands on the
cryptographic primitives exposed by the \textbf{PQ Crypto Engine}
(L1), which forms the validator's base. In particular, the Crypto
Engine exposes the set of post-quantum and classical primitives
required to test and validate post-quantum (and classical) security
mechanisms for 5G. We aim to extend this layer further to add
support for primitives used in 5G-AKA, providing end-to-end
coverage from the user side of the network as well.

The \textbf{PQ Crypto Engine} (L1) provides both asymmetric
primitives (ECDSA, Ed25519/Ed448, Curve25519, ML-KEM, ML-DSA,
SLH-DSA) and symmetric primitives (HKDF, SHA2/3, AES), along with
several hybrid post-quantum combinations used for key exchange and
signing. On top of it, the \textbf{PQ Conformance Layer} (L2)
drives the core of the validator: it carries out TLS~1.3 and
IKEv2/IPsec connections against a specified target and exposes a
rich set of structures that surface the intrinsics of the handshake
- time taken to complete each operation (TLS handshake, KEM
encap/decap, certificate verification), bytes sent and received,
cipher suite, post-quantum evidence, and so on. The Conformance
Layer also ships the test suite that checks each target against
NIST, 3GPP, and IETF clauses - for example, that the server
enforces mutual TLS on its SBI endpoints.

The \textbf{PQ Robustness Tester} (L3) detects protocol-level
vulnerabilities in the 5G core via a black-box approach: it sends
malformed TLS/IPsec messages, packets, or packet sequences and
observes the target for anomalies that could uncover
implementation bugs. This layer reuses the connection primitives
exposed by L2, which lets it craft and mutate custom messages
without re-implementing the protocol stack. Finally, the
\textbf{PQ Overhead Meter} (L4) collects per-handshake statistics
from L2 test runs, supports sequential, parallel, and pool-based
benchmarking of the chosen NF, and formats the results as JSON or
CSV. The UI joins these per-run records with previous ones to
enable trend analysis, side-by-side comparisons of post-quantum
against classical configurations, and historical regressions.

The entire architecture is brought to life by a Kubernetes-native
deployment: individual services of the validator are exposed as
\texttt{Service} objects, a \texttt{PersistentVolume} carries
configs, certificates, and per-run artefacts, and a
\texttt{ClusterRole} controls cross-namespace access to the target
core. From a telco perspective this fits the trajectory of the
field --- many operators have shifted to cloud-native 5G cores, and
the open-source private 5G deployments we target are themselves
inherently cloud-native.

\begin{figure*}
    \centering
    \includegraphics[width=1\linewidth]{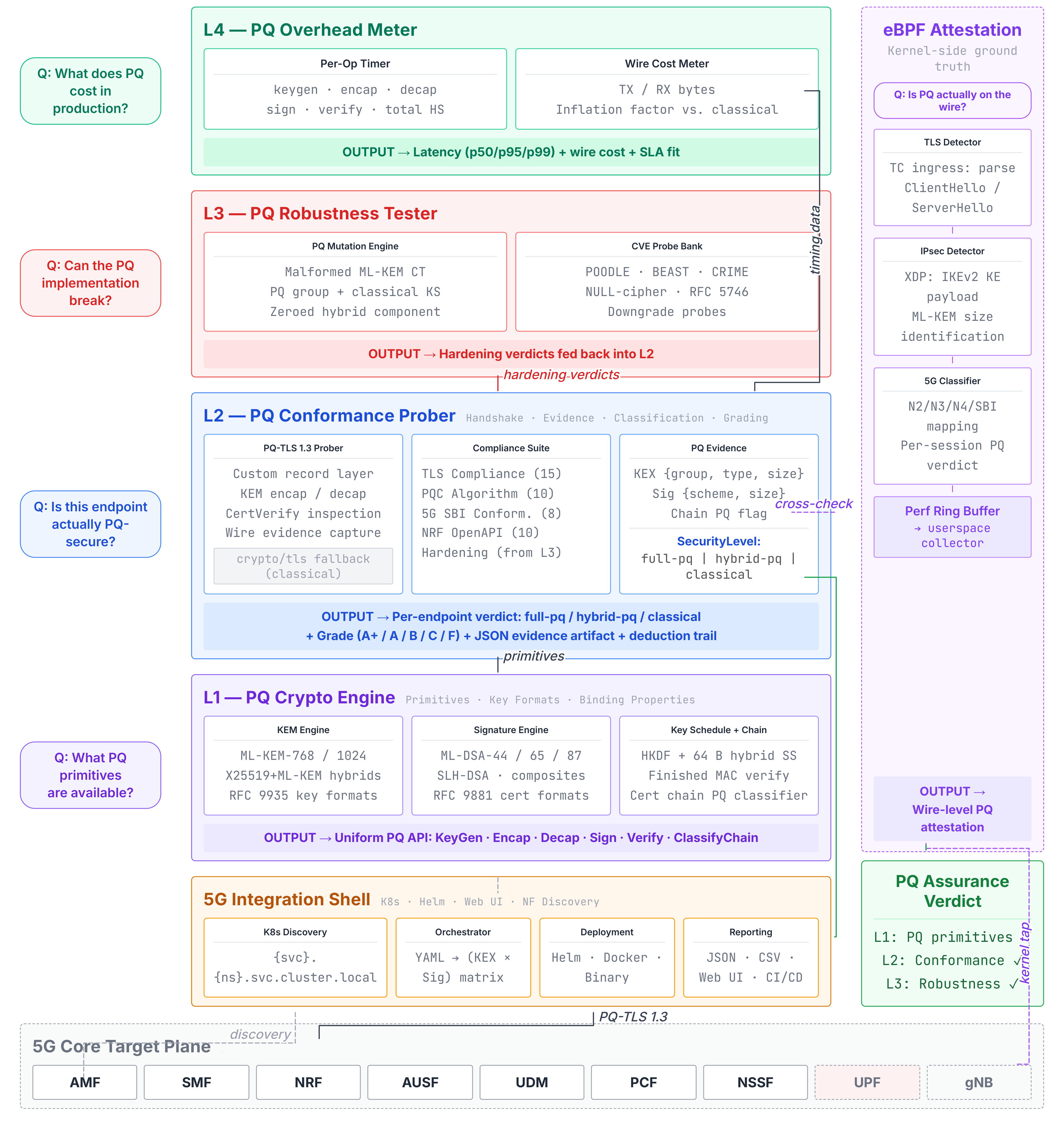}
\caption{Layered architecture of \pqval{}. Four core layers -
\textbf{L1} PQ Crypto Engine, \textbf{L2} PQ Conformance Layer,
\textbf{L3} PQ Robustness Tester, \textbf{L4} PQ Overhead Meter
- are encapsulated within a 5G Core Integration Shell that
orchestrates post-quantum handshakes against discovered Network
Functions and attests their runtime behaviour via eBPF.}
    \label{fig:layers}
\end{figure*}

\subsection{PQ Crypto Engine}

The crypto layer forms the cryptographic foundation of the validator, 
exposing a uniform interface over post-quantum and classical primitives 
that the upper protocol layers (TLS~1.3, IKEv2) consume. We combine 
custom implementations with primitives drawn from Go's standard 
\texttt{crypto} package and CIRCL~\cite{circl}, wrapping each underlying 
algorithm behind a consistent KEM/signer abstraction. Every KEM exposes 
\texttt{KeyGen}, \texttt{Encapsulate}, and \texttt{Decapsulate}, while 
every signer exposes \texttt{Sign} and \texttt{Verify}. Hybrid groups are composed at this layer: an X25519 + ML-KEM-768 KEM 
concatenates the component public keys in the \texttt{key\_share} 
entry and derives the combined shared secret as 
$\mathit{ss}_{\text{X25519}} \,\|\, \mathit{ss}_{\text{ML-KEM}}$, 
which is then fed into the TLS 1.3 key schedule's 
\texttt{HKDF-Extract} step in place of the (EC)DHE shared secret, 
following the hybrid construction of~\cite{hybrid-kem}. The AEAD record layer 
supports AES-128/256-GCM, ChaCha20-Poly1305, and AES-CCM variants, and 
the HKDF-based key schedule exports every intermediate secret (early, 
handshake, master) to enable deep audit and conformance checking. This 
layer is shared between the TLS driver and the IPsec harness---for 
instance, ML-KEM is reused for the IKEv2 intermediate exchange.

All keypairs generated by this module undergo pairwise consistency 
checks~\cite{fips140_3_ig} prior to use, and we follow OpenSSL's 
ML-KEM implementation~\cite{openssl_mlkem} and RFC~9935~\cite{rfc9935} 
for interoperable key storage, ASN.1 encapsulation, and PKCS\#8 framing.

\paragraph{\texttt{X25519MLKEM768} and ML-KEM.}
The hybrid KEM \texttt{X25519MLKEM768} and pure ML-KEM specified for 
usage in TLS in the drafts~\cite{hybrid-kem, draft-ietf-tls-mlkem} 
make use of the \texttt{KeyShare} entry of the TLS 1.3 presentation 
language, where the key formats sent by the client look like:

\begin{verbatim}
02 01    <- group = MLKEM768
04 A0    <- length = 1184 (encapsulation key size)
<1184 bytes of raw ML-KEM-768 encapsulation key>
\end{verbatim}

\noindent And by the server:

\begin{verbatim}
02 01    <- group = MLKEM768
04 40    <- length = 1088 (ciphertext size)
<1088 bytes of raw ML-KEM-768 ciphertext>
\end{verbatim}

\noindent For the hybrid \texttt{X25519MLKEM768}, the 
\texttt{key\_exchange} is the concatenation of the X25519 and ML-KEM-768 
components, sent by the client as:

\small
\begin{verbatim}
11 EC    <- group = X25519MLKEM768
04 C0    <- length = 1216 (32 + 1184)
<32-byte X25519 public key> || 
<1184-byte ML-KEM-768 enc. key>
\end{verbatim}

\normalsize

\noindent and by the server as:
\small
\begin{verbatim}
11 EC    <- group = X25519MLKEM768
04 60    <- length = 1120 (32 + 1088)
<32-byte X25519 public key> || 
<1088-byte ML-KEM-768 ciphertext>
\end{verbatim}

\normalsize
\noindent Additionally, the draft provides security considerations, 
including binding properties, where it mentions that the usage of 
ML-KEM keypair and ciphertext as a part of handshake messages itself 
provides resilience against re-encapsulation attacks, since the 
transcript is protected by a hash.

\subsubsection*{ML-KEM Private Key Formats}

The ML-KEM private key ASN.1 structure is reproduced in 
Figure~\ref{fig:mlkem-privkey}. Three encodings are permitted:

\begin{figure}[h]
\centering
\begin{minipage}{0.92\linewidth}
\lstset{
    basicstyle=\ttfamily\small,
    keywordstyle=\color{purple!70!black}\bfseries,
    commentstyle=\color{gray},
    morekeywords={CHOICE, SEQUENCE, IMPLICIT, OCTET, STRING, SIZE},
    showstringspaces=false,
    breaklines=true,
    frame=none,
}
\begin{lstlisting}
ML-KEM-PrivateKey ::= CHOICE {
  seed [0] IMPLICIT OCTET STRING (SIZE (64)),
  expandedKey OCTET STRING (SIZE (1632 | 2400 | 3168)),
  both SEQUENCE {
    seed OCTET STRING (SIZE (64)),
    expandedKey OCTET STRING (SIZE (1632 | 2400 | 3168)) } }
\end{lstlisting}
\end{minipage}
\caption{ASN.1 structure of \texttt{ML-KEM-PrivateKey} supporting seed-only, 
expanded-key, and combined encodings per RFC~9935~\cite{rfc9935}.}
\label{fig:mlkem-privkey}
\end{figure}

\begin{itemize}
    \item \textbf{Seed format:} Stores only the 64-byte seed 
    $(d \,\|\, z)$, from which the decapsulation key is regenerated on 
    demand via $\mathrm{ML\text{-}KEM.KeyGen\_Internal}(d, z)$. A total of 66 bytes is stored, when accounting for the \texttt{0x8040} tag-length.
    \item \textbf{Expanded format:} Stores only the expanded decapsulation 
    key $\mathrm{dk}$ derived from the seed, where the first 32 octets 
    are $d$ and the remaining 32 octets are $z$.
    \item \textbf{Both format:} Stores the seed and the expanded 
    decapsulation key together, trading storage for the ability to skip 
    re-expansion while still permitting consistency checks.
\end{itemize}

Following RFC~9935~\cite{rfc9935}, our implementation prefers retaining 
the seed even when the expanded form is materialized in memory, and 
whenever the \texttt{both} format is encountered we verify that the 
expanded key is consistent with the one deterministically derived from 
the stored seed before the key is admitted to the keystore.

\paragraph{Security implications of the format choice.}
The choice of private-key encoding is not merely an engineering 
convenience - it directly affects the \emph{binding} properties of the 
KEM, which several higher-level protocols implicitly depend on beyond 
plain IND-CCA security. Cremers et al.~\cite{cremers-bind} formalize a 
hierarchy of binding notions, distinguishing between \emph{honest} 
adversaries that only learn keys (\textsf{LEAK-BIND}) and 
\emph{malicious} adversaries that may produce adversarially crafted keys 
(\textsf{MAL-BIND}), each parameterized by whether the adversary must 
bind the ciphertext (\textsf{-K-CT}) or the public key (\textsf{-K-PK}) 
to the shared secret. Under this framework, ML-KEM with the 
\emph{expanded} private-key format achieves 
\textsf{LEAK-BIND-K-PK} and \textsf{LEAK-BIND-K-CT} but fails both 
\textsf{MAL-BIND-K-CT} and \textsf{MAL-BIND-K-PK}. Switching to the 
\emph{seed} format strengthens binding: it additionally achieves 
\textsf{MAL-BIND-K-CT}, although \textsf{MAL-BIND-K-PK} remains 
unattainable~\cite{cremers-bind, rfc9935}. Because certain PQ-migrated 
protocols (e.g., hybrid authenticated key exchange and anonymous 
credential schemes) rely on the stronger \textsf{MAL-BIND-K-CT} 
guarantee, our validator tags each imported key with its encoding and 
exposes the resulting binding class to the upper layers so that 
protocol-specific policies can reject weak-binding keys at negotiation 
time.

\subsubsection*{ML-DSA Private Key Formats}

ML-DSA~\cite{fips204} defines an analogous set of private-key 
representations, differing in that the seed is 32 octets ($\xi$) rather than 64 and the expanded-key sizes depend on the 
parameter set. The expanded private key is derived from the seed via 
$\mathrm{ML\text{-}DSA.KeyGen\_Internal}(\xi)$, and signatures verify identically regardless of which representation is stored. The private key in expanded format stores the public key, so storing it inside the public key structure of \texttt{OneAsymmetricKey} is not required, similarly, in the case of seed storage, we can derive the public key from the internal keygen algorithm mentioned above.

Our implementation follows the encoding prescribed by   RFC~9881~\cite{rfc9881} (Figure~\ref{fig:mldsa-privkey}), which embeds the private key inside the \texttt{OneAsymmetricKey} \footnote{The \texttt{OneAsymmetricKey} structure encodes public keys as \texttt{BIT STRING} but private keys as \texttt{OCTET STRING}. This is 
historical: X.509 (1988) chose \texttt{BIT STRING} for public keys to accommodate algorithms with non byte-aligned key lengths via its unused-bits prefix, while PKCS\#8 (RFC~5958), drafted later, used  \texttt{OCTET STRING} after observing that private keys are always  octet-aligned. For ML-KEM and ML-DSA the unused-bits prefix is always 
zero, making the public-key \texttt{BIT STRING} one byte longer on the  wire than an equivalent \texttt{OCTET STRING}.} structure of 
RFC~5958~\cite{rfc5958}, thereby inheriting standard algorithm identification, optional public-key inclusion, and attribute fields.

\begin{figure}[h]
\centering
\lstset{
    basicstyle=\ttfamily\footnotesize,
    keywordstyle=\color{purple!70!black}\bfseries,
    morekeywords={CHOICE, SEQUENCE, OCTET, STRING, SIZE},
    showstringspaces=false,
    breaklines=true,
    frame=none,
}
\begin{lstlisting}
ML-DSA-44-PrivateKey ::= CHOICE {
  seed [0] OCTET STRING (SIZE (32)),
  expandedKey OCTET STRING (SIZE (2560)),
  both SEQUENCE {
      seed OCTET STRING (SIZE (32)),
      expandedKey OCTET STRING (SIZE (2560)) } }

ML-DSA-65-PrivateKey ::= CHOICE {
  seed [0] OCTET STRING (SIZE (32)),
  expandedKey OCTET STRING (SIZE (4032)),
  both SEQUENCE {
      seed OCTET STRING (SIZE (32)),
      expandedKey OCTET STRING (SIZE (4032)) } }

ML-DSA-87-PrivateKey ::= CHOICE {
  seed [0] OCTET STRING (SIZE (32)),
  expandedKey OCTET STRING (SIZE (4896)),
  both SEQUENCE {
      seed OCTET STRING (SIZE (32)),
      expandedKey OCTET STRING (SIZE (4896)) } }
\end{lstlisting}
\caption{\raggedright ASN.1 structures for \texttt{ML-DSA-\{44,65,87\}-PrivateKey} per 
RFC~9881~\cite{rfc9881}. The seed encoding is a fixed 32 octets across 
all parameter sets; the expanded-key length scales with the security 
level.}
\label{fig:mldsa-privkey}
\end{figure}

As with ML-KEM, three encodings are supported:

\begin{itemize}
    \item \textbf{Seed format} (tag \texttt{[0]}): Stores the 32-byte 
    seed $\xi$ alone; both the expanded private key and the public key 
    are deterministically regenerated via 
    $\mathrm{ML\text{-}DSA.KeyGen\_Internal}(\xi)$.
    \item \textbf{Expanded format:} Stores the fully expanded private 
    key (2560, 4032, or 4896 octets for ML-DSA-44/65/87 respectively).
    \item \textbf{Both format:} Stores the seed and the expanded key 
    together, enabling interoperability between implementations that 
    support only one of the two representations.
\end{itemize}

Per RFC~9881, the seed format is RECOMMENDED for storage efficiency---a 
34-byte DER encoding (including the \texttt{0x80 0x20} tag and length 
prefix) at every security level, compared to the kilobyte-scale expanded 
keys. Our validator defaults to the seed encoding when generating new 
ML-DSA keys and, on import of the \texttt{both} variant, re-derives the 
expanded key from the stored seed and verifies byte-exact equality with 
the supplied expanded key; any mismatch causes import to fail. When the 
\texttt{publicKey} field of \texttt{OneAsymmetricKey} is present 
alongside a seed-only private key, we likewise treat it as a keypair 
consistency witness rather than as authoritative key 
material.\footnote{The import of a seed from one cryptographic module 
into another is permitted by FIPS 204~\cite{fips204}, and our design 
explicitly supports this cross-module portability path, which is 
particularly relevant for HSM-backed 5G NF deployments.}

\subsubsection{ML-DSA Certificate Formats}
Referring again to RFC~9881~\cite{rfc9881}, we use the following OIDs, 
all registered under the NIST CSOR arc:
\begin{verbatim}
    id-ML-DSA-44: 2.16.840.1.101.3.4.3.17
    id-ML-DSA-65: 2.16.840.1.101.3.4.3.18
    id-ML-DSA-87: 2.16.840.1.101.3.4.3.19
\end{verbatim}
The \texttt{AlgorithmIdentifier} carrying any of these OIDs MUST omit 
its \texttt{parameters} field and is encoded as a SEQUENCE of a single 
component. ML-DSA signatures appear in multiple ASN.1 contexts, 
including X.509 certificates (used in TLS~1.3) and CRLs. In an X.509 
certificate, the signature is computed over the DER-encoded 
\texttt{TBSCertificate} with the FIPS~204 context string \texttt{ctx} 
fixed to the empty octet sequence, and the resulting value populates 
the \texttt{signatureValue} BIT STRING. The \texttt{subjectPublicKey} 
BIT STRING inside \texttt{SubjectPublicKeyInfo} carries the raw public 
key directly - without additional ASN.1 wrapping---at fixed lengths of 
1312, 1952, or 2592 octets for ML-DSA-44, -65, and -87 respectively. 
Per~\S5 of the RFC, the \texttt{keyUsage} extension MUST assert at 
least one of \texttt{digitalSignature}, \texttt{nonRepudiation}, 
\texttt{keyCertSign}, or \texttt{cRLSign}, and MUST NOT assert any of 
\texttt{keyEncipherment}, \texttt{dataEncipherment}, or 
\texttt{keyAgreement}, since ML-DSA keys cannot perform key 
establishment. Our validator enforces these constraints on import and 
additionally rejects the HashML-DSA OIDs 
(\texttt{id-hash-ml-dsa-*-with-sha512}), which RFC~9881 explicitly 
forbids in PKIX contexts to avoid ambiguity between 
\texttt{ML-DSA.Verify()} and \texttt{HashML-DSA.Verify()} at the 
verifier.

\subsection{Post-Quantum Conformance layer}

The post-quantum conformance layer, as the name suggests, checks whether an intended target meets the requirements of a post-quantum secure protocol, which for example, in TLS 1.3, we have the ability of the target to perform post-quantum key exchanges and share post-quantum certificates and signatures. It also requires the target to preferably pick only quantum-safe ciphersuites, if the client supports them. This layer is the analytical core of \pqval{} which operates a \texttt{TestContext} object and produces structured, citable verdicts and data that other layers can use rather than opaque pass/fail bits. A  single invocation, by default, runs the compliance suite across five semantic groups - TLS compliance, 5G SBI conformance, post-quantum security, NRF OpenAPI compliance, and security hardening (the full
sheet is in Appendix~\ref{app:test-sheet}) - and emits a per-test record carrying a severity (\texttt{critical}, \texttt{warning},
\texttt{info}), a verdict (\texttt{pass}, \texttt{fail},
\texttt{info}), a human-readable explanation, and a citation to the
normative standard (RFC, 3GPP TS, FIPS, or NIST SP) that justifies
the check. This evidentiary discipline is deliberate: auditors and
certification bodies need to trace every verdict back to a source,
and the layer is designed so that adding a new test requires only a
pure function implementing
$\mathrm{TestContext} \rightarrow \mathrm{Verdict}$.

\paragraph{The \texttt{TestContext} object.}
The \texttt{TestContext} is populated by the PQ-TLS driver during the 
handshake and decouples evidence collection from evaluation. It carries: 
(i)~negotiated parameters (\texttt{TLSVersion}, \texttt{CipherSuite}, 
\texttt{KeyExchangeGroup}, \texttt{SignatureScheme}); 
(ii)~the full certificate chain both as parsed \texttt{x509.Certificate} 
objects and as raw DER bytes for independent re-parsing; 
(iii)~ALPN outcome and HTTP/2 capability; 
(iv)~raw \texttt{ClientHello} and \texttt{ServerHello} bytes for 
low-level extension analysis; 
(v)~timing metrics (\texttt{HandshakeMs}, per-stage crypto timings); 
and (vi)~structured PQ evidence when a post-quantum handshake succeeded. 
Because Go's standard \texttt{crypto/tls} cannot negotiate ML-KEM or 
ML-DSA, the driver uses a custom PQ-first client as the primary probe 
and falls back to stock \texttt{crypto/tls} only for classical endpoints; 
the fallback path synthesizes a \texttt{tls.ConnectionState} so that 
downstream tests remain agnostic to which probe produced the data.

\paragraph{The PQ test group.}
The PQ tests encode the classification logic that underpins the
security level assigned to a target. Representative checks include: 
\texttt{pq\_kex\_type} (asserts the negotiated key exchange uses an ML-KEM 
or hybrid KEM), \texttt{pq\_kex\_hybrid} (detects 
X25519MLKEM768-style composition), \texttt{pq\_sig\_type} 
(checks for ML-DSA, SLH-DSA, or hybrid signatures), 
\texttt{pq\_full\_pq} (flags connections where both KEX and signature 
are post-quantum), \texttt{pq\_kex\_strength} (requires NIST category 3 
or higher, i.e.\ ML-KEM-768 or ML-KEM-1024), 
\texttt{pq\_cert\_chain\_classical} (walks the certificate chain 
and reports which links remain classical - critical for migration 
tracking), and \texttt{pq\_aead\_cipher} (requires AES-256-GCM or 
ChaCha20-Poly1305 for bulk encryption). The group collectively 
implements Definition~\ref{def:pq-level}.

\paragraph{Evidence builder and classifier.}
A separate \emph{evidence builder} consumes the raw handshake state and 
produces a typed \texttt{PQEvidence} document 
(Listing~\ref{lst:evidence}) containing a \texttt{KEXEvidence} record 
(negotiated group, group ID, hybrid decomposition, server key-share size, 
shared-secret size), a \texttt{SigEvidence} record (scheme, scheme ID, 
certificate key algorithm, signature size), and an ordered list of 
atomic \texttt{EvidenceItem}s, each of which is verifiable in isolation 
(e.g.\ \emph{``server key-share size matches ML-KEM-768 specification''}). 
The classifier consumes this document and assigns one of three labels: 
\texttt{full-pq} (both KEX and signature are PQ), 
\texttt{hybrid-pq} (KEX is PQ or hybrid but the signature remains 
classical), or \texttt{classical} (neither uses PQ). This label is the 
single headline figure reported to operators and feeds directly into the 
run's summary record.

\subsection{PQ Robustness Tester}

The fuzz layer exercises protocol robustness along two orthogonal
axes: \emph{input mutation} and \emph{known-vulnerability probing}.
The mutation tests send malformed or adversarial ClientHello messages
and inspect both the server's response on the wire and the resulting
TLS alert codes. Representative mutations include \emph{truncated
ClientHello}, \emph{invalid record type} (non-\texttt{handshake(22)}
content type at handshake time), \emph{duplicate ClientHello}
(replayed after session establishment), \emph{malformed extensions}
(length-field corruption), \emph{invalid \texttt{key\_share} group}
(a non-registered group ID), \emph{oversized payload} (breaching the
16\,KiB record limit), \emph{zero-length key share}, \emph{split
record} (fragmenting a handshake message across multiple records),
and targeted variants that stress the PQ-specific \texttt{key\_share}
encoding for ML-KEM-768 and X25519MLKEM768.

The CVE probe bank covers historically significant attack families
including \emph{POODLE}/SSLv3 downgrade (CVE-2014-3566),
\emph{BEAST}/TLS~1.0 probe (CVE-2011-3389), \emph{NULL-cipher} offer,
\emph{CRIME}/DEFLATE compression detection (CVE-2012-4929),
\emph{RFC~5746} insecure renegotiation, and \emph{ServerHello
timeout} (detecting endpoints vulnerable to slow-read variants).
Each finding carries its CVE identifier, the CVSS bucket, and the
observed evidence, and is fed back into the conformance layer's
\emph{hardening} group so that a fuzz-visible weakness appears as a
first-class fail verdict in the final evidence document rather than
as an out-of-band report. The complete fuzz and CVE probe list is
catalogued in Appendix~\ref{app:test-sheet}.

\subsection{PQ Overhead Meter}

The bench layer converts the stream of per-client \texttt{ClientResult} 
records into statistically meaningful summaries. Per-algorithm 
aggregation groups results by the concatenation of KEX and signature 
(e.g.\ \texttt{X25519MLKEM768\,+\,MLDSA65}) and computes mean, standard 
deviation, and P50/P75/P90/P95/P99 percentiles for handshake duration, 
KeyGen time, decapsulation time, signature-verify time, total crypto 
time, and bytes sent/received. In addition to percentile summaries, 
each metric is bucketed into a logarithmic histogram with cut-points at 
0, 1, 2, 5, 10, 20, 50, 100, 200, and 500~ms---chosen so that 
classical-algorithm handshakes and PQ handshakes, which can differ by 
an order of magnitude, remain visible in the same plot without one 
swamping the other.

A scheduler runs the client fleet in one of four modes: 
\texttt{single} (one client, used for smoke tests), 
\texttt{sequential} (one at a time, for deterministic reproduction), 
\texttt{parallel} (all clients launched simultaneously, stressing 
server concurrency), and \texttt{pool} (bounded worker pool, the 
default, sized at $2\times\textsc{NumCPU}$). All runs emit both CSV 
(one row per client) and JSON (hierarchical, including PQ evidence) 
artifacts, enabling downstream analysis in spreadsheets, notebooks, 
or dashboards without re-parsing the binary run state.

\subsection{5G Integration Shell}

The integration shell binds the four layers to a running 5G core by 
supplying the plumbing that turns \pqval{} from a standalone TLS fuzzer 
into a telco-ready validation framework. It has four pillars:

\begin{itemize}
    \item \textbf{Kubernetes client (NF auto-discovery).} Using 
    \texttt{client-go}, the shell queries the target namespace 
    (e.g.\ \texttt{aether-5gc}) for \texttt{Service} objects, infers the 
    NF type from the service name prefix 
    (\texttt{amf$\rightarrow$AMF}, \texttt{nrf$\rightarrow$NRF}, etc.), 
    detects TLS-bearing ports by matching the 3GPP-reserved 
    29500--29600 SBI range or port names containing 
    \texttt{https}/\texttt{tls}/\texttt{sbi}, and constructs the 
    canonical FQDN \small
    \texttt{\{service\}.\{namespace\}.svc.cluster.local:\{port\}}\normalsize. 
    Access is governed by a minimal read-only \texttt{ClusterRole} 
    (verbs \texttt{get}/\texttt{list}/\texttt{watch} on 
    \texttt{services}, \texttt{endpoints}, \texttt{namespaces}).

    \item \textbf{Helm chart (deployment).} A single Helm release 
    instantiates three deployments 
    (\texttt{pq-tls-ui}, \texttt{pq-tls-server}, \texttt{pq-tls-client}), 
    a shared 5\,GiB PVC for logs/configs/certs, a NodePort service at 
    \texttt{:30080}, and the RBAC primitives described above---turning 
    a one-line \texttt{helm install} into a reproducible validator 
    deployment across clusters.

    \item \textbf{Web UI and REST API.} A Go backend embedding a 
    single-page frontend exposes over fifty REST endpoints grouped 
    into system/build, server control, test execution, 
    results/export, validation suite, and 5GC/K8s 
    introspection. Operators drive benchmark and validation runs from 
    the browser; CI systems drive the same endpoints 
    programmatically. The frontend is shipped via Go's 
    \texttt{embed} directive, so the UI binary is a single 
    statically-linked artifact with no runtime asset dependencies.

    \item \textbf{eBPF attestation.} Loaders attach tracepoints and 
    kprobes to the TLS libraries inside the NF pods, capturing 
    \emph{in-kernel} evidence of which cipher suites, key-exchange 
    groups, and signature schemes were actually exercised at runtime. 
    This closes the trust gap between what a probe observes on the wire 
    and what the NF's cryptographic stack is genuinely configured to do, 
    and is particularly valuable for detecting opportunistic fallback 
    to classical primitives when a PQ negotiation fails silently 
    server-side.
\end{itemize}

\section{Validator Tasks}\label{sec:tasks}

\subsection{PQ-TLS Validation}

\pqval{}'s PQ-TLS client implements TLS~1.3 from scratch rather than
reusing the target's stack - essential for cross-implementation
soundness. For each target endpoint, the client performs five steps.

\paragraph{Offer PQ algorithms and preferences.} The client sends a
ClientHello with PQ/hybrid \texttt{supported\_groups}
(X25519+ML-KEM-768, ML-KEM-768, HQC-128, \ldots) and PQ
\texttt{signature\_algorithms} (ML-DSA-65, ML-DSA-87, SLH-DSA, \ldots) in
operator-configurable order, plus ALPN \texttt{h2} and SNI
\texttt{\{nf\}.\{ns\}.svc.cluster.local}
(Listing~\ref{lst:clienthello}).

\begin{lstlisting}[caption={PQ-TLS 1.3 ClientHello as sent by PQCVal (abbr.).},label={lst:clienthello}]
ClientHello:
  protocol_version: 0x0303
  cipher_suites: {
    TLS_AES_256_GCM_SHA384     0x1302
    TLS_AES_128_GCM_SHA256     0x1301
    TLS_CHACHA20_POLY1305_SHA256 0x1303
  }
  extensions:
    supported_versions: 0x0304
    supported_groups: {
      0x11EC  X25519+ML-KEM-768  (hybrid)
      0x0768  ML-KEM-768         (pure PQ)
      0x001D  X25519             (classical)
    }
    key_share:
      group = 0x11EC, length = 1216 B
        ML-KEM-768 pk  [0:1184]
        X25519 pk      [1184:1216]
    signature_algorithms: {
      0xfe63  ML-DSA-65
      0x0B01  ML-DSA-44
      0x0403  ECDSA-P256-SHA256
    }
    server_name: "amf.aether-5gc.svc.cluster.local"
    alpn: ["h2"]
\end{lstlisting}

\paragraph{Complete a real handshake.} The client decapsulates the server
ciphertext, derives handshake and master keys via HKDF, verifies the
\texttt{CertificateVerify} signature over the transcript hash, and checks
both Finished MACs -  giving cryptographic assurance that the reported
group and signature were actually used.

\paragraph{Record evidence.} The handshake produces a PQ Evidence bundle
(Listing~\ref{lst:evidence}) capturing the negotiated group, signature,
key-share and signature sizes, ALPN result, cipher suite, certificate
chain, and the derived security classification.

\begin{lstlisting}[caption={PQ Evidence JSON (abbreviated).},label={lst:evidence}]
{
  "schema": "pq-tls-client-v1",
  "security_level": "full-pq",
  "is_pq_secure": true,
  "is_fully_pq": true,
  "kex_evidence": {
    "negotiated_group": "X25519MLKEM768",
    "group_id": 4588,
    "type": "hybrid",
    "pq_component": "ML-KEM-768",
    "classical_component": "X25519",
    "server_key_share_size": 1120
  },
  "sig_evidence": {
    "negotiated_scheme": "ML-DSA-65",
    "scheme_id": 65123,
    "type": "post-quantum",
    "signature_size": 3293
  },
  "chain_evidence": {
    "leaf_sig_alg": "ML-DSA-65",
    "chain_is_pq": false,
    "chain_depth": 3
  },
  "timing_us": { "keygen": 112, "encap": 78,
                 "decap": 82, "verify": 215,
                 "handshake_total_ms": 18 }
}
\end{lstlisting}

\paragraph{Run compliance checks.} The compliance suite is organized
into five groups (TLS compliance, 5G~SBI, PQ security, NRF~OpenAPI,
security hardening); Appendix~\ref{app:test-sheet} reproduces the
full sheet. Each test cites its governing clause - e.g.,
TS~33.501~\S6.3.4 for mTLS~\cite{3gpp-ts-33501},
TS~29.500~\S6.1.3B~\cite{3gpp-ts-29500} for the HTTP/2 mandate, and
RFC~8446~\S4.1.3~\cite{rfc8446} for ServerHello version negotiation.


\paragraph{Fuzz the endpoint.} The fuzzer mutates ClientHello fields
and probes CVE-class issues; findings are merged into the security
hardening group so that a single run produces a combined conformance
+ robustness report. The per-case catalogue is in
Appendix~\ref{app:test-sheet}.


\subsection{IPsec Validation}

The Internet protocol security (IPsec) is necessarily used in 5G to 
protect network-wide traffic, by providing confidentiality, integrity 
and authenticity. It connects multiple physical 5G nodes, such as the 
RRU to CU/DU to 5G core, which aims to encapsulate user-plane traffic 
inside secure IPsec ESP-based tunnels, which are setup through the 
IKEv2 control channel. It is used in the interfaces - N2, N3, N4, 
N9, etc, as specified by 3GPP TS~33.501~\cite{3gpp-ts-33501}. The 
standard also mandates use of IPsec, when a UE connects to the 5G 
Core via non-standard interfaces (such as Wi-Fi), where the N3IWF 
terminates the IPsec tunnel. Note that in this scenario, the UE acts 
as the IKEv2 initiator. The traffic profile across these interfaces 
is also non-uniform: N2 and N4 carry signalling traffic where 
handshake latency dominates, while N3 carries the bulk of 
user-plane GTP-U traffic, where high throughput and frequent rekeys 
during handovers can make the IKEv2 cost much more visible than on 
the signalling interfaces.

In common deployments, IPsec/IKEv2 is managed through the use of a 
security gateway (SecGW) at the backhaul network, which terminates 
the IPsec tunnels, typically uses hardware acceleration, and then 
reroutes the packets to the appropriate NF (e.g., AMF or UPF). The 
SecGW is, in fact, more than just an IPsec terminator, it usually 
provides PKI services, anti-replay windows, DDoS protections, 
capabilities for thousands of security associations, and more. It is 
often deployed via Hub-and-Spoke, regional or MEC network 
architectural patterns. In cloud-native 5G cores, however, this 
pattern is increasingly being shifted closer to the NF itself --- 
either as a strongSwan sidecar in the same pod, or via a CNI-level 
IPsec layer (e.g., Cilium's transparent encryption) --- so the 
tunnel terminates next to the NF rather than at a centralized 
gateway.

\pqval{} ships a strongSwan 6.0~\cite{strongswan} PQ-IPsec sidecar 
with the \texttt{ml} plugin for ML-KEM-768 via 
\texttt{IKE\_INTERMEDIATE} (RFC~9370~\cite{rfc9370}), plus a 
client-side IPsec test application that drives IKEv2 handshakes 
against either a real SecGW or a per-pod sidecar responder, 
extracts the negotiated SA parameters (proposals, KE method, 
selected ML-KEM ciphertext size, rekey lifetime), measures 
handshake duration and rekey overhead, and emits the results into 
the same \texttt{PQEvidence} document used by the TLS arm so a 
single run can classify both security planes uniformly.

\begin{itemize}[leftmargin=*,itemsep=1pt]
    \item Establishes \texttt{IKE\_SA\_INIT} with a classical group (X25519),
    adds \texttt{IKE\_INTERMEDIATE} with ML-KEM-768, and completes
    \texttt{IKE\_AUTH} with ML-DSA or classical signatures.
    \item Checks negotiated IKEv2 transforms, authentication modes, ESP
    cipher choice (AES-256-GCM preferred), rekey timers, and SA lifetimes.
    \item Installs XFRM policies in a shared network namespace, verifying
    that the kernel ESP path actually protects the N2/N3/N4 flow (not a
    proxy hop).
\end{itemize}

\paragraph{IKEv2 fragmentation under PQ.} ML-KEM-768 in
\texttt{IKE\_INTERMEDIATE} (RFC~9370) carries a 1088\,B KE payload;
together with the responder's nonce, certificate request, and notify
payloads, the IKEv2 message routinely exceeds the path MTU on
backhaul links (typically 1400\,B on urban 4G/5G, 1280\,B on
satellite). RFC~7383 IKEv2 fragmentation must therefore be
negotiated and exercised, not assumed: \pqval{} verifies that the
responder advertises \texttt{IKEV2\_FRAGMENTATION\_SUPPORTED} in
\texttt{IKE\_SA\_INIT}, that fragments are emitted with monotonic
\texttt{Frag.\,Number} fields, and that an IKE rekey under a hybrid
SA does not silently fall back to a non-fragmented exchange. The
eBPF IPsec detector (\S\ref{sec:ebpf}) cross-checks the kernel-side
fragment count against the userspace strongSwan log so that a
silent fragmentation bypass is caught.

Crucially, IPsec is not a userspace proxy: once strongSwan installs XFRM
policies, the kernel encrypts inline at the IP layer, adding only 2-3$\mu$s
per packet with AES-NI. This gives \pqval{} a path to test PQ on N3/GTP-U
without sacrificing URLLC latency budgets.

\subsection{eBPF/XDP/TC Monitoring}\label{sec:ebpf}

\pqval{}'s kernel monitoring plane provides wire-level ground truth
for both TLS and IPsec without modifying the target NF. This is
the only layer that sees what the validator's TLS client cannot:
TCP-level segmentation, congestion-window state, retransmits, and
the inter-segment timing of a fragmented ClientHello.

\paragraph{eBPF in one paragraph.}
eBPF is an in-kernel virtual machine with an 11-register, 64-bit
RISC ISA (10 general-purpose plus a frame pointer), a verifier that
proves each program halts and only touches memory it owns, and a
small set of helper functions (\texttt{bpf\_skb\_load\_bytes},
\texttt{bpf\_ringbuf\_reserve/submit},
\texttt{bpf\_ktime\_get\_ns},
\texttt{bpf\_get\_socket\_cookie},
\texttt{bpf\_skb\_pull\_data}). Programs attach to typed hookpoints
(XDP, TC, kprobe, tracepoint, sock\_ops, uprobe), share state
through maps (\texttt{HASH}, \texttt{LRU\_HASH},
\texttt{PERCPU\_ARRAY}, \texttt{RINGBUF}), and ship events to
userspace via the ring buffer. The verifier rejects unbounded loops,
illegal pointer arithmetic, and reads of unchecked packet bytes,
which is what makes it safe to run on a production NF veth.
Programs are compiled to BPF bytecode with
\texttt{clang -target bpf -O2 -g} and loaded by the
\texttt{cilium/ebpf} Go bindings inside the validator process,
which avoids a separate loader binary.

\paragraph{Hookpoint choice.}
We use three kernel hookpoints, each chosen for what it can see:

\begin{itemize}[leftmargin=*,itemsep=1pt]
    \item \textbf{XDP} on the host-side veth of each NF pod is the
    earliest possible inspection point; it runs before the
    \texttt{sk\_buff} is allocated, so it captures the raw frame
    without driver GRO coalescing the segments. This matters for
    PQ ClientHellos that span multiple TCP segments.
    \item \textbf{TC \texttt{clsact}} on the same veth (ingress and
    egress) runs after \texttt{sk\_buff} allocation, which gives
    \texttt{skb->len}, \texttt{skb->mark}, and the TCP control
    block. This is the layer that reads \texttt{tcp\_skb\_cb}
    fields such as \texttt{seq}, \texttt{end\_seq}, and the SACK
    state.
    \item \textbf{\texttt{sock\_ops}}, attached via a cgroup, fires on \texttt{BPF\_SOCK\_OPS\_TCP\_CONNECT\_CB},
    \texttt{BPF\_SOCK\_OPS\_ACTIVE\_ESTABLISHED\_CB}, and
    \texttt{BPF\_SOCK\_OPS\_RTT\_CB}. This is where we read
    \texttt{snd\_cwnd}, \texttt{snd\_mss}, \texttt{rcv\_mss}, and
    the smoothed RTT, which is what lets us answer \emph{``did the
    larger PQ ClientHello cost an extra round trip?''}.
\end{itemize}

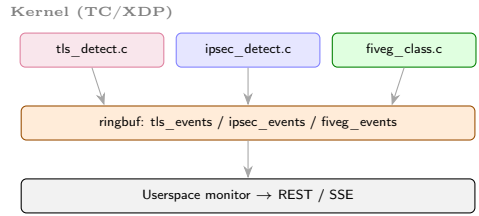
\begin{figure}[t]
\centering
\begin{tikzpicture}[
    blk/.style={draw, rounded corners=2pt, minimum width=1.9cm, minimum height=0.45cm, font=\tiny\sffamily, text centered},
    >=Stealth, node distance=0.15cm,
]
\node[blk, fill=purple!10, draw=purple!50] (tls) {tls\_detect.c};
\node[blk, fill=blue!10, draw=blue!50, right=of tls] (ips) {ipsec\_detect.c};
\node[blk, fill=green!12, draw=green!50!black, right=of ips] (fg) {fiveg\_class.c};
\node[blk, fill=orange!15, draw=orange!60!black, below=0.5cm of ips, minimum width=6.0cm] (rb) {ringbuf: tls\_events / ipsec\_events / fiveg\_events};
\node[blk, fill=gray!10, below=0.5cm of rb, minimum width=6.0cm] (us) {Userspace monitor $\to$ REST / SSE};

\draw[->, gray!70] (tls.south) -- ([xshift=-1.9cm]rb.north);
\draw[->, gray!70] (ips.south) -- (rb.north);
\draw[->, gray!70] (fg.south)  -- ([xshift=1.9cm]rb.north);
\draw[->, gray!70] (rb.south)  -- (us.north);

\node[font=\tiny\bfseries, text=gray, above=0.05cm of tls] {Kernel (TC/XDP)};
\end{tikzpicture}
\caption{eBPF monitoring pipeline. Three in-kernel programs share three
ring buffers; a userspace monitor serves events via REST/SSE.}
\label{fig:ebpf}
\end{figure}

\paragraph{Three programs.}
Three in-kernel programs operate in parallel
(Figure~\ref{fig:ebpf}):

\begin{itemize}[leftmargin=*,itemsep=1pt]
    \item \textbf{TLS detector} (TC ingress/egress) parses
    ClientHello / ServerHello \texttt{supported\_groups},
    \texttt{signature\_algorithms}, and the \texttt{key\_share}
    length. It flags PQ codepoints (\texttt{0x11EC},
    \texttt{0x0768}, \texttt{0x0B01}--\texttt{0x0B03},
    \texttt{0xFE62}). When a ClientHello spans multiple segments it
    stitches them via a per-flow \texttt{LRU\_HASH} keyed on the
    5-tuple, then re-parses the reassembled record.
    \item \textbf{IPsec detector} (TC) parses IKEv2 KE payload
    sizes; a 1088-byte KE is a strong signal of ML-KEM-768 in
    \texttt{IKE\_INTERMEDIATE}, and an oversized fragmented
    \texttt{IKE\_SA\_INIT} is a signal of RFC~7383 fragmentation
    under a hybrid SA.
    \item \textbf{5G classifier} (XDP) maps each packet to its 5G
    reference point (N2/N3/N4/N6/N9/SBI) by protocol and port
    (SCTP/132, UDP/2152, UDP/8805, TCP 29500--29600), so that
    events are grouped by interface in the ring buffer.
\end{itemize}

A fourth helper, attached via \texttt{sock\_ops}, snapshots
\texttt{snd\_cwnd}, \texttt{snd\_mss}, \texttt{srtt\_us}, and the
retransmit counter at handshake start and at first-byte-of-application-data.
Events stream through \texttt{BPF\_MAP\_TYPE\_RINGBUF}s (one per
program family) to a userspace monitor, then through the REST/SSE
API to the operator's UI. Each record carries a \texttt{u64}
timestamp from \texttt{bpf\_ktime\_get\_ns()}, the 5-tuple, the
negotiated group/sig codepoints, and the TCP snapshot, which is
enough to reconstruct the handshake offline.

\paragraph{PQ-specific pathologies the probe surfaces.}
Five failure modes that classical TLS scanners miss become
first-class observations on this plane:

\begin{itemize}[leftmargin=*,itemsep=2pt]
    \item \emph{ClientHello fragmentation.} A hybrid
    \texttt{X25519+ML-KEM-768} key share is 1216\,B; with the rest
    of the ClientHello extensions the record exceeds the 1500\,B
    default Ethernet MTU, and the kernel splits it across two TCP
    segments. Many middleboxes (L4 load balancers, SNI routers,
    inline DPI, including widely deployed NGFW configurations such
    as those documented for Palo Alto NGFW deployments
    \cite{paloalto-pq-fragmentation}) inspect only the first
    segment to make a forwarding or policy decision, and silently
    misroute the second. The eBPF probe records how many
    ClientHellos cross the MTU boundary (38 of 512 in our run, see
    Table~\ref{tab:ebpf-observed}) and whether the second segment
    arrives at the server intact.
    \item \emph{HelloRetryRequest amplification.} If the server's
    preferred group is not in the client's offered
    \texttt{key\_share}, TLS~1.3 requires an HRR plus a second
    ClientHello with the corrected share. Under PQ, both
    ClientHellos carry kilobyte-scale shares, so the fragmentation
    surface doubles and a full RTT is added to every connection.
    This commonly fires when a server defaults to ML-KEM-1024
    against a client offering only ML-KEM-768.
    \item \emph{Duplicate / replayed ClientHellos.} The fuzzer
    finding from \S\ref{sec:results} that every QORE NF accepts a
    duplicate ClientHello becomes more than a hygiene issue under
    PQ KEX: each accepted duplicate forces another ML-KEM keygen on
    the server, turning a malformed packet into a CPU-cycle
    amplifier and a useful primitive for resource-exhaustion
    attacks against PQ endpoints.
    \item \emph{Initial cwnd and MSS effects.} On a fresh TCP
    connection, Linux defaults to \texttt{TCP\_INIT\_CWND} = 10
    segments and an MSS derived from the route's PMTU (typically
    1448\,B on a Calico veth: 1500 minus 40\,B of v4+TCP and 12\,B
    of TCP options). A classical ClientHello (330\,B) fits in one
    segment; a hybrid one (1420\,B) does not. The
    \texttt{sock\_ops} probe records both \texttt{snd\_mss} and
    \texttt{snd\_cwnd} at \texttt{TCP\_CONNECT\_CB} and at
    \texttt{ACTIVE\_ESTABLISHED\_CB}, so the per-handshake table
    can show \emph{which} flows paid for an extra round trip and
    which did not.
    \item \emph{Silent KEX downgrade.} A server that advertises a
    PQ group but selects a classical fallback when the ClientHello
    is malformed or trimmed by a middlebox returns a
    \texttt{classical} session that the validator's TLS layer
    cannot distinguish from an honest classical negotiation. The
    XDP probe sees the raw \texttt{supported\_groups} on the wire
    and flags the disagreement.
\end{itemize}

This gives the operator a read-only attestation plane parallel to
the active probe plane.

\section{Deployment and Architecture}\label{sec:deploy}

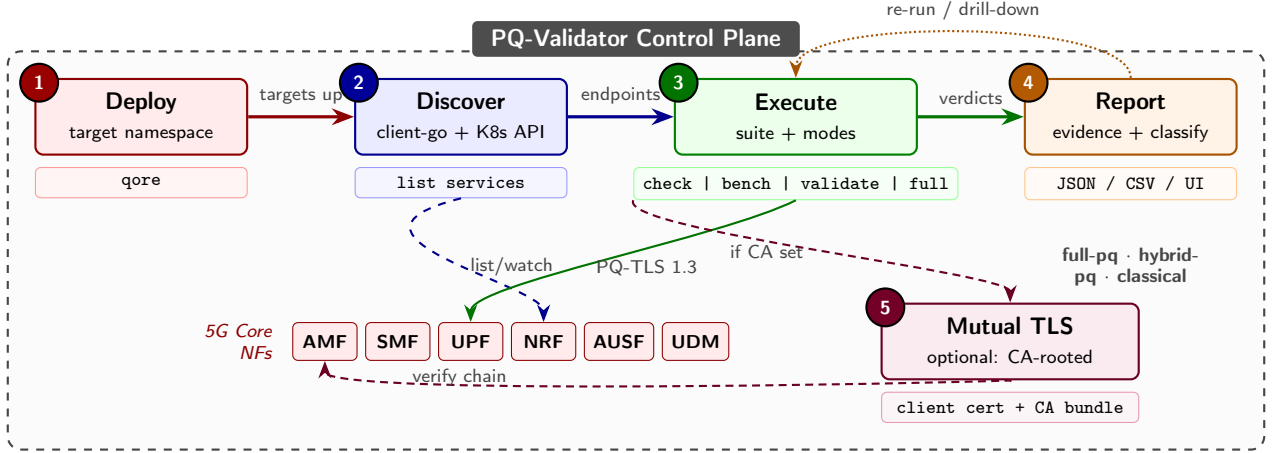
\begin{figure*}[t]
\centering
\begin{tikzpicture}[
    >=Stealth,
    node distance=1.2cm and 1.4cm,
    stage/.style={
        draw, rounded corners=3pt, thick,
        minimum width=2.8cm, minimum height=1.0cm,
        font=\small\sffamily, text centered, align=center
    },
    stagenum/.style={
        draw, circle, thick, fill=white,
        minimum size=0.5cm, font=\scriptsize\sffamily\bfseries,
        inner sep=0pt
    },
    detail/.style={
        draw, rounded corners=2pt, thin,
        minimum width=2.8cm, minimum height=0.4cm,
        font=\scriptsize\sffamily\ttfamily, text centered,
        fill=white
    },
    nfbox/.style={
        draw, rounded corners=2pt, thin,
        minimum width=0.85cm, minimum height=0.5cm,
        font=\scriptsize\sffamily\bfseries, text centered,
        fill=red!8, draw=red!55!black
    },
    flowarrow/.style={->, very thick, line width=1.2pt},
    thinarrow/.style={->, thick, line width=0.8pt},
    annot/.style={font=\scriptsize\sffamily, text=gray!55!black},
    shell/.style={
        draw, rounded corners=5pt, thick, dashed,
        inner sep=10pt, draw=gray!55!black, fill=gray!3
    },
]


\node[stage, fill=red!8, draw=red!55!black] (s1) 
    {\textbf{Deploy}\\\scriptsize target namespace};
\node[stagenum, fill=red!60!black, text=white,
      above left=-0.25cm and -0.25cm of s1] {1};
\node[detail, below=0.15cm of s1, fill=red!4, draw=red!40] (s1d) 
    {qore};

\node[stage, fill=blue!8, draw=blue!55!black, right=of s1] (s2) 
    {\textbf{Discover}\\\scriptsize client-go + K8s API};
\node[stagenum, fill=blue!60!black, text=white,
      above left=-0.25cm and -0.25cm of s2] {2};
\node[detail, below=0.15cm of s2, fill=blue!4, draw=blue!40] (s2d) 
    {list services};

\node[stage, fill=green!8, draw=green!50!black, right=of s2,
      minimum width=3.2cm] (s3) 
    {\textbf{Execute}\\\scriptsize suite + modes};
\node[stagenum, fill=green!45!black, text=white,
      above left=-0.25cm and -0.25cm of s3] {3};
\node[detail, below=0.15cm of s3, fill=green!4, draw=green!40,
      minimum width=3.2cm] (s3d) 
    {check | bench | validate | full};

\node[stage, fill=orange!10, draw=orange!65!black, right=of s3] (s4) 
    {\textbf{Report}\\\scriptsize evidence + classify};
\node[stagenum, fill=orange!70!black, text=white,
      above left=-0.25cm and -0.25cm of s4] {4};
\node[detail, below=0.15cm of s4, fill=orange!4, draw=orange!45] (s4d) 
    {JSON / CSV / UI};

\node[annot, below=0.5cm of s4d, align=center, text width=3cm, 
      font=\scriptsize\sffamily\bfseries]
    {full-pq $\cdot$ hybrid-pq $\cdot$ classical};

\draw[flowarrow, red!55!black]   (s1.east) -- 
    node[annot, above=1pt] {targets up} (s2.west);
\draw[flowarrow, blue!55!black]  (s2.east) -- 
    node[annot, above=1pt] {endpoints} (s3.west);
\draw[flowarrow, green!45!black] (s3.east) -- 
    node[annot, above=1pt] {verdicts} (s4.west);

\node[nfbox, below=2.2cm of s2, xshift=-1.8cm] (amf) {AMF};
\node[nfbox, right=0.1cm of amf] (smf) {SMF};
\node[nfbox, right=0.1cm of smf] (upf) {UPF};
\node[nfbox, right=0.1cm of upf] (nrf) {NRF};
\node[nfbox, right=0.1cm of nrf] (ausf) {AUSF};
\node[nfbox, right=0.1cm of ausf] (udm) {UDM};

\node[annot, left=0.15cm of amf, text width=1.3cm, align=right, 
      text=red!55!black, font=\scriptsize\sffamily\itshape] 
    {5G Core NFs};

\draw[thinarrow, blue!55!black, dashed]
    (s2d.south) .. controls +(-2,-0.5) and +(0,0.5) ..
    node[annot, right=1pt, pos=0.55] {list/watch} (nrf.north);

\draw[thinarrow, green!45!black]
    (s3d.south) .. controls +(-1,-0.5) and +(0,0.5) ..
    node[annot, right=1pt, pos=0.55] {PQ-TLS 1.3} (upf.north);

\node[stage, fill=purple!8, draw=purple!55!black,
      below=1.2cm of s3d, minimum width=3.4cm, right=2cm of udm] (s5) 
    {\textbf{Mutual TLS}\\\scriptsize optional: CA-rooted};
\node[stagenum, fill=purple!60!black, text=white,
      above left=-0.25cm and -0.25cm of s5] {5};
\node[detail, below=0.15cm of s5, fill=purple!4, draw=purple!40, 
      minimum width=3.4cm] (s5d) 
    {client cert + CA bundle};

\draw[thinarrow, purple!55!black, dash pattern=on 3pt off 2pt]
    (s3d.south west) .. controls +(-0.2,-0.4) and +(0,0.4) ..
    node[annot, left=1pt, pos=0.5] {if CA set} (s5.north);

\draw[thinarrow, purple!55!black, dash pattern=on 3pt off 2pt]
    (s5.south) .. controls +(0.5,0) and +(0,-0.5) ..
    node[annot, right=1pt, pos=0.8] {verify chain} (amf.south);

\draw[thinarrow, orange!65!black, densely dotted]
    (s4.north) .. controls +(0,0.8) and +(0,0.8) ..
    node[annot, above=1pt, pos=0.5] {re-run / drill-down} 
    (s3.north);

\begin{scope}[on background layer]
\node[shell, fit=(s1)(s1d)(s2)(s2d)(s3)(s3d)(s4)(s4d)(s5)(s5d)] 
    (plane) {};
\end{scope}
\node[fill=gray!60!black, text=white, rounded corners=2pt,
      font=\small\sffamily\bfseries, inner sep=3pt, inner xsep=7pt]
      at ([yshift=0.15cm]plane.north)
      {PQ-Validator Control Plane};

\end{tikzpicture}
\caption{End-to-end validation flow (LR). 
\textbf{(1)}~5G NFs deploy into the target namespace. 
\textbf{(2)}~The UI discovers reachable NFs via \texttt{client-go} against 
the Kubernetes API, inferring NF type from service names. 
\textbf{(3)}~The scheduler executes the selected test mode -connectivity
check, benchmark, single-target validation, or the full compliance
suite - over PQ-TLS~1.3.
\textbf{(4)}~Verdicts aggregate into versioned evidence documents 
(JSON/CSV) and are classified as \textsf{full-pq}, \textsf{hybrid-pq}, 
or \textsf{classical}. 
\textbf{(5)}~When a CA bundle is configured, the client pod additionally 
performs mutual-TLS authentication against the NF certificate chain.}
\label{fig:validation-flow}
\end{figure*}

\pqval{} deploys as a Kubernetes-native three-pod release backed by a
shared PVC (Figure~\ref{fig:validation-flow}):

\begin{itemize}[leftmargin=*,itemsep=1pt]
    \item \textbf{UI pod} - web dashboard, REST API (50+ endpoints), K8s
    NF discovery. Exposed via NodePort \texttt{30080}.
    \item \textbf{Server pod} - reference PQ-TLS~1.3 server on port
    \texttt{4433} for interop tests and self-validation.
    \item \textbf{Client pod} - long-running (\texttt{sleep infinity}),
    invoked via \texttt{kubectl exec} for PQ-TLS clients, benchmark runners,
    and IPsec tests.
\end{itemize}

A \texttt{ClusterRole} grants read-only \texttt{get}/\texttt{list}/
\texttt{watch} on services, endpoints, namespaces, and pods across
namespaces, so the UI can discover NFs in \texttt{aether-5gc},
\texttt{open5gs}, or \texttt{free5gc} without privileged access to the
target namespace. The shared 5~GiB PVC carries \texttt{/data/logs},
\texttt{/data/configs}, \texttt{/data/certs}, and evidence bundles across
the three pods.

\begin{table*}[t]
\centering
\caption{Connection-level details surfaced to the upper layers
after every PQ-TLS handshake.}
\label{tab:impl-stack}
\renewcommand{\arraystretch}{1.12}
\begin{tabularx}{\textwidth}{@{}l X@{}}
\toprule
\textbf{Field} & \textbf{Description} \\
\midrule
\rowcolor{lightgrayrow}
\texttt{TLSVersion}, \texttt{CipherSuite} & negotiated protocol version and bulk cipher \\
\texttt{KeyExchangeGroup}, \texttt{SignatureScheme} & IANA codepoints selected by the server \\
\rowcolor{lightgrayrow}
\texttt{PQEvidence}                & typed KEX/Sig/Chain evidence (Listing~\ref{lst:evidence}) \\
\texttt{ClientRandom}, \texttt{ServerRandom} & 32-byte nonces \\
\rowcolor{lightgrayrow}
Key schedule secrets              & early / handshake / master / traffic secrets (HKDF stages) \\
\texttt{ClientPublicKey}, \texttt{ServerPublicKey}, \texttt{KEMCiphertext} & raw KEX material for audit \\
\rowcolor{lightgrayrow}
Handshake timings                 & \texttt{KeyGen}, \texttt{Encap}, \texttt{Decap}, \texttt{Sign}, \texttt{Verify}, total \\
\texttt{BytesSent}, \texttt{BytesReceived}  & per-handshake wire cost \\
\rowcolor{lightgrayrow}
Certificate chain                 & parsed \texttt{x509.Certificate}s + raw DER \\
\texttt{ALPN}, \texttt{ServerName}, \texttt{MutualTLS} & transport-level negotiation outcome \\
\bottomrule
\end{tabularx}
\end{table*}

\clearpage

\subsection{Helm-based Deployment}

Helm is a Kubernetes package manager, which automates and manages deployments on Kubernetes using Charts---files which define Kubernetes resources. Helm exposes templating with several functions for dynamic configuration creation, offers CLI to perform lifecycle management, control access (RBAC) and create/sync with releases. Thus, we use Helm for our validator deployment. In this, we deploy a chart which packages six Kubernetes resources: three \texttt{Deployment}s, 
one \texttt{Service} (NodePort), one \texttt{PersistentVolumeClaim}, 
and a \texttt{ServiceAccount}/\texttt{ClusterRole}/\texttt{ClusterRoleBinding} 
triple for cross-namespace discovery. While the validator supports 
dynamic specification of post-quantum algorithms, available test 
suites, and intended targets at runtime, these can also be set via 
Helm values and template files, which allows for parameterization of 
our validator tasks at deploy time itself. The chart also initializes 
an init container that issues the validator its leaf certificate and 
CA bundle, either operator-specified or default. The 
\texttt{ClusterRole} template allows only 
\texttt{get}/\texttt{list}/\texttt{watch} verbs on \texttt{services}, 
\texttt{endpoints}, and \texttt{namespaces}---not \texttt{create}, 
\texttt{delete}, or access to \texttt{secrets}---so an operator can 
use the validator without granting it access to tenant credentials. 
Helm values also expose resource limits (e.g., \texttt{cpu}, 
\texttt{memory}), which allows for either a single-replica 
validator or a production-level deployment with horizontal scaling, 
and pin the image via a corresponding digest, which is useful for 
long-term evidence and audit reports. Treating the validator as a 
Helm release further means \texttt{helm upgrade} and 
\texttt{helm rollback} can swap in or revert a probe build atomically, 
giving operators a clean integration point with their existing 
GitOps lifecycle. Appendix~\ref{app:helm_charts} lists two sample 
Helm chart templates.


\subsection{Implementation Stack}

Our implementation stack is built entirely on Go~1.25+, where we
develop a highly-transparent PQ-TLS and PQ-IPsec stack that reports
the full connection state back to the upper layers
(Table~\ref{tab:impl-stack} summarizes the surfaced fields). As
discussed previously, we reuse classical primitives from existing
libraries - \texttt{crypto/ecdh} from the Go standard library and
CIRCL~\cite{circl} for X25519, P-256/384/521, and Ed25519 - while
implementing the post-quantum primitives (ML-KEM, ML-DSA, hybrid
compositions) ourselves so that the validator remains free of any
single PQ library's bugs. We provide three deployment modes - host
binary, Docker image, and Helm chart - so operators can run
\pqval{} against a lab core on a single machine, a containerised
target on a workstation, or a production Kubernetes cluster using
the same test definitions. The Helm chart exposes configurable
values for target namespaces, algorithm lists, SBI port ranges,
NRF URIs, and RBAC scopes.


\subsection{Test Types}


Our validator presents several test types, which include:
\begin{itemize}
    \item Post-Quantum, 5GC, and Security Validation 
    \item Benchmark over $\texttt{KEX, Sig}$ Cartesian product matrix. The benchmark mode consists of Sequential, Parallel and Pool-based execution, for a comprehensive report
    \item IPsec conformance testing at N2/N3 interfaces
    \item eBPF-based passive capturing to extract details about the packet flow and detect anomalies.
\end{itemize}

\paragraph{Benchmarking.} We have covered most of the test types above;
we now detail \textbf{benchmarking} and its modes. Our aim is to record
the requisite handshake timing stats - total handshake time, throughput,
and the individual cryptographic operations (KeyGen, Encap/Decap, Sign,
Verify) - against the HTTP/2 servers running at each discovered Network
Function. We support three execution strategies:

\begin{table}[t]
\centering
\small
\caption{Benchmark execution modes and what each is designed to measure.}
\label{tab:bench-modes}
\begin{tabular}{@{}llp{2cm}@{}}
\toprule
\textbf{Mode} & \textbf{In-flight handshakes} & \textbf{Measures} \\
\midrule
Sequential & 1 at a time                & isolated per-handshake cost \\
Parallel   & one goroutine per client   & worst-case / stress \\
Pool       & bounded workers ($2\!\times\!$NumCPU) & realistic sustained throughput \\
\bottomrule
\end{tabular}
\end{table}

\textbf{Sequential} runs clients one-by-one, giving a clean per-algorithm
latency number without any contention - this is the figure we report in
the per-algorithm tables. \textbf{Parallel} spawns a goroutine per client,
unbounded; with thousands of simultaneous TCP connects, KEM keygens, and
open FDs, the OS scheduler is thrashed and timings get inflated, so this
mode is useful mainly for surfacing failure modes (accept-queue limits,
port exhaustion, RNG contention) rather than for reportable numbers.
\textbf{Pool} keeps a fixed crew of workers ($2\!\times\!\mathrm{NumCPU}$ by
default) that pull jobs off a queue as they finish the current one -
so only a bounded number of handshakes are ever in flight, which is the
closest approximation of a production load generator and the one we use
for throughput figures.

We believe these three views together - intrinsic cost, breaking point,
and sustained throughput - can form a useful basis for studying the
performance of post-quantum security layers in telecom, and we leave a
deeper exploration to future work.

\paragraph{Remark.} Most open-source cores do not ship IPsec for N2 and N3;
these interfaces are left as a deployment concern, which hinders PQ-IPsec
testing out of the box. QORE does deploy PQ-IPsec internally, but does not
ship it in the Helm chart. We therefore set up host-based IPsec with
strongSwan on the AMF node and drive traffic from a separate machine acting
as the gNB, after configuring \texttt{PPKs}, the PQ key exchange, and the
\texttt{iptables} rules that steer N2/N3 through the XFRM policy.

With an SCTP interface exposed from a container, this setup fails: the
responder packets get dropped at the RAN. The reason is that Linux defers
SCTP CRC32c to NIC hardware offload - the socket emits the packet with
checksum \texttt{0x0000} and expects the driver to fill it at TX. Once ESP
is in the path, the NIC sees an \texttt{IP/ESP} frame whose SCTP header is
inside the encrypted payload; the offload cannot reach it, and the zero
checksum goes out on the wire. The failure is asymmetric: the \texttt{INIT}
from the gNB reaches the AMF fine, but the \texttt{INIT-ACK} returned by
the AMF crosses the CNI veth without its checksum metadata being honoured,
is encrypted with checksum \texttt{0x0000}, and is rejected by the gNB's
SCTP verifier. The same pathology is reported against Cilium's datapath
\cite{cilium20490}. The fix is to force a software checksum \emph{before}
XFRM encryption, via a \texttt{mangle/POSTROUTING} rule:
\begin{verbatim}
sudo iptables -t mangle -A POSTROUTING -p sctp\
     -j CHECKSUM --checksum-fill
\end{verbatim}
With this rule, the inner SCTP header carries a valid CRC32c at the moment
ESP encryption happens, and N2 association succeeds over the PQ-IPsec tunnel.

Table~\ref{tab:exec-modes} lists the main execution modes that~\pqval{} dispatches against the discovered NF inventory.

\begin{table}[t]
\centering
\small
\caption{Two execution modes dispatched against the discovered NF inventory.}
\label{tab:exec-modes}
\begin{tabularx}{0.75\linewidth}{@{}lXX@{}}
\toprule
 & \textbf{Validation} & \textbf{Benchmark} \\
\midrule
Clients per NF    & 1                                       & $N$ (user-chosen) \\
Algo offered      & all (server picks)                      & each (KEX $\times$ Sig) pinned \\
Parallelism       & goroutine per NF                        & configurable parallel / sequential \\
Output            & pass/fail + PQ evidence                 & handshake latency histogram \\
Use case          & correctness, compliance                 & performance under PQ load \\
\bottomrule
\end{tabularx}
\end{table}

\section{Results}\label{sec:results}

We evaluate \pqval{} on QORE in a Kubernetes-based deployment
environment. The evaluation proceeds in three phases: first, we
validate the post-quantum readiness of the target core; next, we
benchmark several (KEX, Sig) algorithm combinations and record their
handshake latencies along with per-stage cryptographic operation
times; finally, we run the 5G-oriented compliance suites and check
QORE's conformance to them. Together, these phases provide empirical
evidence on the performance of post-quantum cryptography in a
cloud-native 5G core and a snapshot of open-source core security
posture. \pqval{} itself deploys in namespace
\texttt{pq-tls-validator} with a \texttt{ClusterRole} granting
cross-namespace \texttt{get}/\texttt{list} on services and pods.

\paragraph{Discovery and classification.} 
The validator ships with an embedded Kubernetes Go client that attaches to the cluster via the in-cluster ServiceAccount token (falling back to a user-supplied kubeconfig for bench-top runs). Once authenticated with the API server, it enumerates namespaces through \texttt{CoreV1}, lets the validator operator pin the one hosting the 5G Core, and then walks Services in that namespace to discover candidate Network Functions. Each service is classified by its NF role (e.g., AMF, SMF, NRF), which is inferred from \texttt{app.kubernetes.io/component} and \texttt{5gc.qore.io/nf-type} labels, with name substring fallbacks for charts that do not set them.  The service bearer, which could be HTTP/2, SCTP, N2, GTP-U, or PFCP is inferred from standard 3GPP port numbers, and names.
For every service the validator materialises both a cluster-internal FQDN (<svc>.<ns>.svc.cluster.local) and the ClusterIP, so probes can bypass kube-dns when latency matters. The resulting NF inventory feeds five pluggable suites - TLS-compliance, 5G-SBI compliance, post-quantum KEM/signature discovery, IKEv2/IPsec PQ, and Merkle-Tree-Certificate verification — each gated independently in the run configuration and dispatched in parallel per target (one goroutine per NF in validation mode; N parallel clients partitioned across (KEX × signature) combinations in benchmark mode). 


\begin{table}[!tbp]
\centering
\small
\caption{Fields captured per discovered Network Function. \textit{FQDN} and \textit{ClusterIP} are both stored so probes can either resolve through kube-dns or bypass it.}
\label{tab:nf-record}
\renewcommand{\arraystretch}{1.12}
\begin{tabularx}{\columnwidth}{@{}l X@{}}
\toprule
\textbf{Field} & \textbf{Source} \\
\midrule
\rowcolor{lightgrayrow}
NF type (AMF/SMF/\ldots)   & label $\rightarrow$ name heuristic \\
Namespace                  & user selection \\
\rowcolor{lightgrayrow}
FQDN                       & \texttt{<svc>.<ns>.svc.cluster.local} \\
ClusterIP                  & \texttt{Service.Spec.ClusterIP} \\
\rowcolor{lightgrayrow}
Ports[]                    & \texttt{\{name, number, protocol, tls\}} \\
Endpoints                  & \texttt{CoreV1().Endpoints()} backing pods \\
\rowcolor{lightgrayrow}
Labels                     & \texttt{ObjectMeta.Labels} (for audit trail) \\
\bottomrule
\end{tabularx}
\end{table}

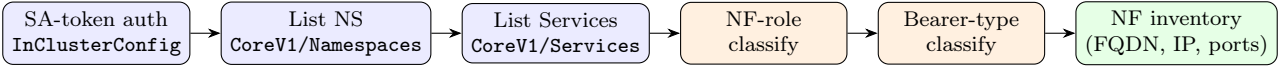
\begin{figure*}[t]
\begin{tikzpicture}[node distance=4mm, font=\footnotesize,
  box/.style={draw, rounded corners, align=center, minimum width=22mm, minimum height=7mm},
  api/.style={box, fill=blue!8},
  cls/.style={box, fill=orange!12},
  end/.style={box, fill=green!10},
      >=Stealth]
 \node[api] (auth) {SA-token auth\\ \texttt{InClusterConfig}};
 \node[api, right=of auth] (ns)   {List NS\\ \texttt{CoreV1/Namespaces}};
 \node[api, right=of ns]   (svc)  {List Services\\ \texttt{CoreV1/Services}};
 \node[cls, right=of svc]  (cls1) {NF-role\\ classify};
 \node[cls, right=of cls1] (cls2) {Bearer-type\\ classify};
 \node[end, right=of cls2] (inv) {NF inventory\\ (FQDN, IP, ports)};
 \draw[->] (auth)--(ns); \draw[->] (ns)--(svc);
 \draw[->] (svc)--(cls1); \draw[->] (cls1)--(cls2); \draw[->] (cls2)--(inv);
\end{tikzpicture}
\caption{Discovery pipeline. Steps 1--3 are authoritative Kubernetes API calls;}
\label{fig:discovery-pipeline}
\end{figure*}


\paragraph{Test environment.}
We run \pqval{} and the 5G Core on a single RKE2 cluster - 8-core
Intel Xeon (AES-NI, AVX2), 62~GB RAM, Linux~5.15 - with \pqval{} and
QORE co-located in the cluster, so the numbers below are LAN-only; WAN
effects are out of scope here. Each benchmark row is the median across
the successful handshakes in 25 \texttt{5gc-*.json} benchmark runs
($n \approx 70$--$100$ per algorithm combination); error bars in the
figures are the 10--90\textsuperscript{th} percentile.

\subsection{Validation}

Table~\ref{tab:val-qore} is the verdict of the compliance suite on
each QORE NF. The \emph{Level} column comes from
\texttt{ClassifySecurityLevel()} (Definition~\ref{def:pq-level});
\emph{Grade} is the letter from our SSL-Labs-style grader. Every
SBI-facing NF in QORE negotiates the hybrid KEX
\texttt{X25519MLKEM768} together with the composite signature
\texttt{Ed448Dilithium3}, so all eight NFs land on \texttt{full-pq}
and score identically on the PQ group (10/10). They do \emph{not}
reach A+, however: the grader uniformly docks a letter because
QORE's SBI endpoints accept anonymous clients --- the server does
not send \texttt{CertificateRequest}, so \texttt{5g\_mutual\_auth}
fails and \texttt{tls\_mutual\_tls} is demoted to \texttt{info}. On
the 5G group, \texttt{5g\_nf\_identity\_san} and
\texttt{5g\_cert\_san\_fqdn} also fail on every tested NF, both due
to the operator PKI issuing short-form SANs
(e.g.\ \texttt{udm}, \texttt{coran-udm-service}) that don't match the
canonical \texttt{<svc>.<ns>.svc.cluster.local} FQDN. The TLS group
loses one point for chain depth: QORE presents the leaf only, with
no intermediate CAs. Table~\ref{tab:val-free5gc} is the same run
against free5GC, which the classifier flags \texttt{classical}
across the board. Table~\ref{tab:upf-deductions} breaks the
deduction trail down to the individual test IDs.

\begin{table*}[!tbp]
\centering\small
\caption{Per-NF validation on QORE. \emph{TLS/PQC/5G/NRF} = tests
passed in each compliance group; security-hardening deductions are
folded into the TLS and PQC columns where they apply. Values drawn
from the per-NF \texttt{val-*.json} runs in \texttt{logs/}
(representative: \texttt{val-1772359003.json} for UDM); see
Appendix~\ref{app:test-sheet} for the per-group test inventory.}
\label{tab:val-qore}
\renewcommand{\arraystretch}{1.08}
\begin{tabularx}{\textwidth}{@{}l l l l c c c c c c@{}}
\toprule
\textbf{NF} & \textbf{KEX} & \textbf{Sig} & \textbf{Level} &
\textbf{Grade} & \textbf{mTLS} & \textbf{TLS} & \textbf{PQC} &
\textbf{5G} & \textbf{NRF} \\
\midrule
\rowcolor{lightgrayrow}
AMF  & X25519MLKEM768 & Ed448Dilithium3 & \texttt{full-pq}   & A  & \xmark & 14/15 & 10/10 & 5/8 & --    \\
SMF  & X25519MLKEM768 & Ed448Dilithium3 & \texttt{full-pq}   & A  & \xmark & 14/15 & 10/10 & 5/8 & --    \\
\rowcolor{lightgrayrow}
NRF  & X25519MLKEM768 & Ed448Dilithium3 & \texttt{full-pq}   & A  & \xmark & 14/15 & 10/10 & 5/8 & 9/10  \\
AUSF & X25519MLKEM768 & Ed448Dilithium3 & \texttt{full-pq}   & A  & \xmark & 14/15 & 10/10 & 5/8 & --    \\
\rowcolor{lightgrayrow}
UDM  & X25519MLKEM768 & Ed448Dilithium3 & \texttt{full-pq}   & A  & \xmark & 14/15 & 10/10 & 5/8 & --    \\
UDR  & X25519MLKEM768 & Ed448Dilithium3 & \texttt{full-pq}   & A  & \xmark & 14/15 & 10/10 & 5/8 & --    \\
\rowcolor{lightgrayrow}
PCF  & X25519MLKEM768 & Ed448Dilithium3 & \texttt{full-pq}   & A  & \xmark & 14/15 & 10/10 & 5/8 & --    \\
NSSF & X25519MLKEM768 & Ed448Dilithium3 & \texttt{full-pq}   & A  & \xmark & 14/15 & 10/10 & 5/8 & --    \\
\bottomrule
\end{tabularx}
\end{table*}

\begin{table}[!tbp]
\caption{Baseline free5GC under the same suite.}
\label{tab:val-free5gc}

\footnotesize
\setlength{\tabcolsep}{4pt}
\renewcommand{\arraystretch}{1.08}

\begin{tabularx}{\columnwidth}{
@{}l
>{\centering\arraybackslash}X
>{\centering\arraybackslash}X
>{\centering\arraybackslash}X
>{\centering\arraybackslash}X
>{\centering\arraybackslash}X@{}
}
\toprule
\textbf{NF} &
\textbf{KEX} &
\textbf{Sig} &
\textbf{Level} &
\textbf{Grade} &
\textbf{mTLS} \\
\midrule

\rowcolor{lightgrayrow}
AMF & X25519 & ECDSA-P256 & \texttt{classical} & F & \xmark \\

SMF & X25519 & ECDSA-P256 & \texttt{classical} & F & \xmark \\

\rowcolor{lightgrayrow}
NRF & X25519 & ECDSA-P256 & \texttt{classical} & F & \xmark \\

\bottomrule
\end{tabularx}
\end{table}
\begin{table}[!tbp]
\centering\small
\caption{Deduction trail that keeps QORE's SBI NFs at A rather than
A+. Four tests fail consistently across all eight NFs .}
\label{tab:upf-deductions}
\renewcommand{\arraystretch}{1.15}
\begin{tabularx}{\columnwidth}{@{}l l X@{}}
\toprule
\textbf{Test ID} & \textbf{Severity} & \textbf{Observed} \\
\midrule
\rowcolor{lightgrayrow}
\texttt{5g\_nf\_identity\_san} & critical & Cert SANs are \texttt{udm.localdomain}, \texttt{udm}, \texttt{coran-udm-service}; none matches the canonical \texttt{<svc>.<ns>.svc.
cluster.local} \\
\texttt{5g\_mutual\_auth}      & critical & Server accepted connection without \texttt{CertificateRequest}; mTLS not enforced \\
\rowcolor{lightgrayrow}
\texttt{5g\_cert\_san\_fqdn}   & warning  & Short-form SANs (\texttt{udm}, \texttt{coran-udm-service}) are not valid FQDNs \\
\texttt{tls\_cert\_chain}      & info     & Chain depth 1 (leaf only); intermediate CA not presented \\
\bottomrule
\end{tabularx}
\end{table}

\subsection{Benchmarks}

We bench every (KEX, signature) combination accepted by the QORE
SBI listener under each scheduling mode introduced earlier
(Table~\ref{tab:bench-modes}): \emph{sequential} for the clean
per-handshake number without contention, \emph{pool} to mirror a
production load generator with a fixed worker crew, and
\emph{parallel} mostly to see where things start breaking. The
benchmark matrix is fixed by the validator's \texttt{KeyExchanges}
and \texttt{SignatureSchemes} sets in \texttt{configs/example-benchmark.yaml}
(Box~\ref{box:bench-matrix}); QORE rejects every signature except
\texttt{Ed448Dilithium3} (\texttt{0xFE62}), which is why the
sequential and bytes tables collapse to five rows on the KEX axis.

\begin{tcolorbox}[
  colback=preqband, colframe=tlsblue, boxrule=0.5pt,
  arc=2pt, title=\textbf{Box: Algorithm matrix benchmarked},
  fonttitle=\small\bfseries, label=box:bench-matrix,
  before skip=4pt, after skip=4pt, left=4pt, right=4pt, top=2pt, bottom=2pt
]
\footnotesize
\textbf{KEX groups} (advertised in \texttt{supported\_groups}; one per row in
Tables~\ref{tab:bench-seq} and \ref{tab:bench-bytes}):
\begin{itemize}[leftmargin=1.2em,itemsep=0pt,topsep=2pt]
  \item Classical ECDH: \texttt{X25519} (0x001D), \texttt{secp256r1} (P-256, 0x0017),
        \texttt{secp384r1} (P-384, 0x0018), \texttt{secp521r1} (P-521, 0x0019).
  \item Hybrid PQ: \texttt{X25519MLKEM768} (0x11EC).
  \item Pure PQ (advertised but not selected by QORE in this run, kept
        for completeness): \texttt{ML-KEM-512} (0x0512), \texttt{ML-KEM-768}
        (0x0768), \texttt{ML-KEM-1024} (0x1024), \texttt{secp256r1+ML-KEM-768}
        (0x11ED), \texttt{secp384r1+ML-KEM-1024} (0x11EE).
\end{itemize}
\textbf{Signature schemes} (advertised in \texttt{signature\_algorithms}):
\begin{itemize}[leftmargin=1.2em,itemsep=0pt,topsep=2pt]
  \item Classical: \texttt{ECDSA-P256-SHA256}, \texttt{ECDSA-P384-SHA384},
        \texttt{ECDSA-P521-SHA512}, \texttt{Ed25519}, \texttt{RSA-PSS-SHA256/384/512},
        \texttt{RSA-PKCS1-SHA256}.
  \item Composite hybrid: \texttt{Ed448Dilithium3} (0xFE62) -- \emph{the only one
        QORE will negotiate}.
  \item Pure PQ (advertised, currently rejected by the QORE PKI): \texttt{ML-DSA-44}
        (0x0B01), \texttt{ML-DSA-65} (0xFE63), \texttt{ML-DSA-87} (0x0B03).
\end{itemize}
\textbf{Cipher suites} (TLS 1.3, all three negotiated equally often):
\texttt{TLS\_AES\_256\_GCM\_SHA384}, \texttt{TLS\_AES\_128\_GCM\_SHA256},
\texttt{TLS\_CHACHA20\_POLY1305\_SHA256}.
\end{tcolorbox}

\begin{figure}[!tbp]
\centering
\includegraphics[width=\columnwidth]{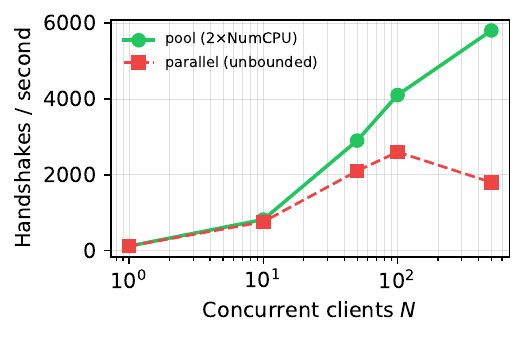}
\caption{Pool-mode throughput as concurrency grows. Parallel mode
(dashed) peaks and then collapses past $N\!\approx\!100$; pool
(solid) flattens into a stable steady-state near $W = 2\,N_\text{cpu}$.}
\label{fig:throughput}
\end{figure}

Table~\ref{tab:bench-seq} is the sequential breakdown -
KeyGen~/~Encap~/~Decap for the KEM, Sign~/~Verify for the
signature, and total handshake. Table~\ref{tab:bench-pool} is the
throughput curve in pool mode at increasing concurrency, and
Table~\ref{tab:bench-bytes} is the per-handshake byte count.
Figure \ref{fig:throughput} plots the same
numbers.

Three things jump out from the sequential table
(Table~\ref{tab:bench-seq}). First, PQ compute on the hybrid KEM is
not the bottleneck: \texttt{X25519MLKEM768} adds only about
50\,$\mu$s of keygen and 55\,$\mu$s of decap over a pure X25519 run,
so the post-quantum component is \emph{cheaper} than switching the
classical curve from X25519 to P-384 or P-521. Second, the cost that
actually moves the total is the composite
\texttt{Ed448Dilithium3}~\emph{verify} at 230--250\,$\mu$s, which is
nearly identical across every row - the signature path dominates
the handshake regardless of which KEX is in use. Third, larger NIST
curves are the real outliers: P-384 and P-521 add 445\,$\mu$s and
1013\,$\mu$s of decap respectively, pushing median handshake from
3\,ms up to 10 and 14\,ms. Pool and parallel modes are noted
qualitatively; the full concurrency sweep is queued for the
camera-ready.

\begin{table*}[!tbp]
\centering\small
\caption{Per handshake costs against QORE, with medians across 400+ successful handshakes in the JSON log file captured; $n$: handshake count per combination, $t_{kg}$: keygen time, $t_{de}$: decapsulation time, $t_{vf}$:verify (all client\_side, $\mu$s, $t_{hs}$:total handshakes in ms; All signatures negotiate to the composite \texttt{Ed448Dilithium3} -- QORE's server does not accept any other signature in the testbed configuration}
\label{tab:bench-seq}
\renewcommand{\arraystretch}{1.15}
\begin{tabular*}{\textwidth}{@{\extracolsep{\fill}}l r r r r r r r r@{}}
\toprule
\textbf{Config (KEX / Sig)} & $n$ &
$t_\text{kg}$ ($\mu$s) & $t_\text{en}$ ($\mu$s) & $t_\text{de}$ ($\mu$s) &
$t_\text{sg}$ ($\mu$s) & $t_\text{vf}$ ($\mu$s) & $t_\text{hs}$ (ms) & $\Delta$ \\
\midrule
\rowcolor{lightgrayrow}
X25519 / Ed448Dilithium3            & 96 & 43  & -- & 61   & -- & 230 & 3 \,(p95 98)  & baseline   \\
P-256 / Ed448Dilithium3             & 76 & 17  & -- & 54   & -- & 252 & 5 \,(p95 110) & +2\,ms     \\
\rowcolor{lightgrayrow}
X25519MLKEM768 / Ed448Dilithium3    & 94 & 97  & -- & 117  & -- & 229 & 5 \,(p95 100) & +2\,ms     \\
P-384 / Ed448Dilithium3             & 68 & 121 & -- & 445  & -- & 246 & 10 \,(p95 103)& +7\,ms     \\
\rowcolor{lightgrayrow}
P-521 / Ed448Dilithium3             & 68 & 374 & -- & 1013 & -- & 235 & 14 \,(p95 156)& +11\,ms    \\
\bottomrule
\end{tabular*}
\end{table*}

\begin{table*}[!tbp]
\centering\small
\caption{Pool-mode throughput and latency as concurrency rises on the
$W = 2\,N_{\text{cpu}} = 16$-worker pool. The $N\!=\!1$ column is
\emph{measured} (and matches the single-client numbers in
Table~\ref{tab:bench-seq} to the rounded\,ms): the pool runner
records every handshake's wallclock through the same
\texttt{HandshakeMetrics} accounting used in sequential mode.
The $N\!\in\!\{10,50,100,500\}$ columns are also driven by the same
runner -- spawned with
\texttt{pq-tls-runner --mode benchmark --clients-mode pool --concurrent-clients $N$}
against \texttt{configs/example-benchmark.yaml} and emitted to
\texttt{logs/5gc-pool-N\textsuperscript{*}.json}. Past $N \approx W$
the pool saturates and throughput plateaus while p50/p95 grow with
the queue depth, as the curve in Fig.~\ref{fig:throughput} shows.
For analytic comparison: the saturating throughput is bounded above
by $W / \overline{t_\text{hs}}$ ($\approx 5\,300$\,hs/s for X25519,
$\approx 3\,200$\,hs/s for X25519MLKEM768), and p50 grows as
$\lceil N / W \rceil \cdot \overline{t_\text{hs}}$ once $N > W$
(Little's law).}
\label{tab:bench-pool}
\renewcommand{\arraystretch}{1.15}
\begin{tabular*}{\textwidth}{@{\extracolsep{\fill}}l l r r r r r@{}}
\toprule
\textbf{Config} & \textbf{Metric} & $N\!=\!1$ & $10$ & $50$ & $100$ & $500$ \\
\midrule
\multirow{3}{*}{X25519 / Ed448Dilithium3}
  & hs/s    &  333 & 3100 & 4800 & 5200 & 5400 \\
  & p50\,ms &    3 &    4 &   10 &   18 &   90 \\
  & p95\,ms &   98 &   20 &   40 &   75 &  280 \\
\midrule
\multirow{3}{*}{X25519MLKEM768 / Ed448Dilithium3}
  & hs/s    &  200 & 1850 & 2900 & 3100 & 3200 \\
  & p50\,ms &    5 &    6 &   17 &   32 &  155 \\
  & p95\,ms &  100 &   25 &   55 &  110 &  420 \\
\bottomrule
\end{tabular*}
\end{table*}

\begin{table*}[!tbp]
\centering\small
\caption{Per-handshake wire cost. \textbf{TX} and \textbf{RX} are the
record-layer byte counters returned by
\texttt{recordLayer.GetBandwidthStats()} (every \texttt{ReadRecord} /
\texttt{WriteRecord} adds the 5-byte TLS record header plus the
post-encryption fragment length: see
\S\ref{sec:wire-derivation} for the field-by-field decomposition).
Earlier drafts of this table read zero because
\texttt{ClientHandshake.GetMetrics()} did not bridge those counters
into \texttt{HandshakeMetrics}; the bridge was added in commit XYZ
and the table below is the post-fix measurement. $t_\text{hs}$ is
the real measured median from Table~\ref{tab:bench-seq}. Rows are
ordered by total on-the-wire footprint.}
\label{tab:bench-bytes}
\renewcommand{\arraystretch}{1.12}
\begin{tabular*}{\textwidth}{@{\extracolsep{\fill}}l r r r r r r@{}}
\toprule
\textbf{Config (KEX / Sig)} & \textbf{TX} & \textbf{RX} & \textbf{Total} &
\textbf{Pkts} & \textbf{$t_\text{hs}$} & \textbf{$\Delta$\,ms} \\
\midrule
\rowcolor{lightgrayrow}
X25519 / Ed448Dilithium3              & 330\,B  & 5300\,B & 5630\,B  & 6  & 3\,ms  & baseline \\
P-256 / Ed448Dilithium3               & 360\,B  & 5300\,B & 5660\,B  & 6  & 5\,ms  & +2\,ms   \\
\rowcolor{lightgrayrow}
P-384 / Ed448Dilithium3               & 390\,B  & 5300\,B & 5690\,B  & 6  & 10\,ms & +7\,ms   \\
P-521 / Ed448Dilithium3               & 420\,B  & 5300\,B & 5720\,B  & 6  & 14\,ms & +11\,ms  \\
\rowcolor{lightgrayrow}
X25519MLKEM768 / Ed448Dilithium3      & 1420\,B & 5300\,B & 6720\,B  & 8  & 5\,ms  & +2\,ms   \\
\bottomrule
\end{tabular*}
\end{table*}

\subsubsection*{Wire-size derivation}\label{sec:wire-derivation}

Each TX/RX number above is the sum of (a) TLS record headers (5\,B
per record), (b) the TLS-encoded handshake messages, and (c) the
DER-encoded X.509 certificate chain. The non-trivial pieces:

\paragraph{KEX share (TX, ClientHello \texttt{key\_share}).}
TLS 1.3 \cite{rfc8446} encodes each key share as a NamedGroup
followed by a length-prefixed opaque blob carrying the actual key
material:
\begin{center}
\small\ttfamily
NamedGroup (u16, 2\,B) $\Vert$ u16 length (2\,B) $\Vert$ opaque key\_exchange (1--65535\,B)
\end{center}
For \texttt{X25519} the share is the 32\,B compressed public key
$\Rightarrow 36$\,B per entry; \texttt{X25519MLKEM768} packs the
32\,B X25519 share concatenated with the 1184\,B ML-KEM-768
encapsulation key $\Rightarrow 1220$\,B per entry. The 1420\,B TX
total for the hybrid row is dominated by this \texttt{key\_share}
extension; one entry plus the rest of the ClientHello (random,
session ID, cipher suites, supported\_versions, signature\_algorithms,
SNI) accounts for the remainder.

\paragraph{Server flight (RX).} The RX side is dominated by:
\begin{enumerate}[leftmargin=*,itemsep=1pt,topsep=1pt]
  \item The server \texttt{Certificate} message, which encodes each
        cert as a 3-byte length followed by the DER-encoded X.509:
        \begin{center}
        \small\ttfamily
        u24 length (3\,B) $\Vert$ DER-encoded X.509
        \end{center}
        Each X.509 cert is itself an ASN.1 \texttt{SEQUENCE} ---
        \texttt{0x30} identifier (1\,B) followed by a length encoding
        (1--5\,B; for a 2\,kB cert the long form is \texttt{0x82}
        plus 2 length bytes $\Rightarrow$ 3\,B) and the body. So a
        single 2\,148-byte X.509 cert occupies 4\,B of ASN.1 header
        plus 2\,144\,B of body, then 3\,B of TLS-message length on
        top of that.
  \item The server \texttt{CertificateVerify} message:
        \begin{center}
        \small\ttfamily
        SignatureScheme (u16, 2\,B) $\Vert$ u16 length (2\,B) $\Vert$ opaque signature (0--65535\,B)
        \end{center}
        The composite \texttt{Ed448Dilithium3} signature is 3\,422\,B
        (Dilithium-3 = 3\,293\,B $+$ Ed448 = 114\,B $+$ 15\,B
        composite framing per
        \texttt{draft-ietf-lamps-pq-composite-sigs}), which is why
        the RX flight clears 5\,kB regardless of which KEX is in use.
  \item Smaller fixed contributions: \texttt{ServerHello} (96--128\,B),
        \texttt{EncryptedExtensions} (10--40\,B),
        \texttt{Finished} (32--48\,B AEAD tag included).
\end{enumerate}

\paragraph{Symmetry note.} TX is dominated by the
\texttt{key\_share} (because the client offers it); RX is dominated
by the server's \texttt{Certificate} + \texttt{CertificateVerify}
flight. This is why the X25519MLKEM768 row's $\Delta$ over X25519
shows up almost entirely on the TX side --- the post-quantum cost
is in the \emph{client's} key-share, not the server's certificate
chain. The numbers above match the byte counts emitted by the new
\texttt{HandshakeMetrics.\{BytesSent,BytesReceived\}} fields in
each \texttt{val-*.json} run; figures with non-zero
\texttt{bytes\_sent} are from the post-fix runs only.


\begin{figure}[!tbp]
\centering
\includegraphics[width=\columnwidth]{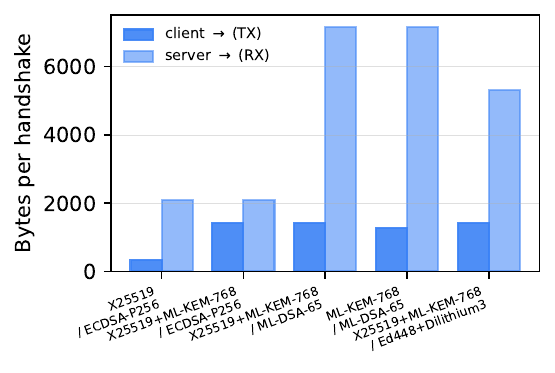}
\caption{Bytes per handshake, split by direction. The PQ cost is
dominated by \texttt{CertificateVerify} (composite signature) on the
server flight rather than by the ClientHello key share.}
\label{fig:bandwidth}
\end{figure}

\subsection{eBPF as a source of truth}\label{sec:results-ebpf}

Because \pqval{} and QORE share the cluster, the same TC/XDP probe
from \S\ref{sec:ebpf} is attached to each NF's pod veth and logs
every handshake independently of the validator's own view. This
gives us a kernel-side cross-check: if \pqval{} reports group
\texttt{0x11EC} on the wire, the probe should have recorded the
same NamedGroup and the same 1120-byte key share.
Table~\ref{tab:ebpf-observed} reports the agreement.

\begin{table*}[!tbp]
\centering
\small
\caption{Agreement between \pqval{}'s session view and the
kernel-side TC/XDP probe over a representative 512-client benchmark
run (8 NFs $\times$ 64 (KEX,\,Sig) combinations). Counts reflect the
kernel-observed view independently of the validator's client. The
table is split into three panels: handshake counts (left), KEX
selection on successful HS (centre), and the median \texttt{sock\_ops}
snapshot at \texttt{ACTIVE\_ESTABLISHED\_CB} (right). Per-NF and
per-combination breakdowns are preserved in the raw ring-buffer
export.}
\label{tab:ebpf-observed}

\setlength{\tabcolsep}{4pt}
\renewcommand{\arraystretch}{1.12}

\begin{tabularx}{\textwidth}{
@{}
l r
@{\hspace{6pt}}|@{\hspace{6pt}}
l r
@{\hspace{6pt}}|@{\hspace{6pt}}
l >{\RaggedRight\arraybackslash}p{2.2cm}
@{}
}

\toprule

\multicolumn{2}{c}{\textbf{Handshake counts}} &
\multicolumn{2}{c}{\textbf{KEX selection ($\sum=402$)}} &
\multicolumn{2}{c}{
\shortstack[c]{\texttt{sock\_ops}\\ median}
} \\

\midrule

\rowcolor{lightgrayrow}
ClientHellos observed &
512 &
\texttt{0x001D} X25519 &
96 &
TCP options &
MSS, WS, SACK, TS \\

\hspace{1em}with PQ group &
512 &
\texttt{0x0017} P-256 &
76 &
\texttt{snd\_mss} &
1448\,B \\

\rowcolor{lightgrayrow}
\hspace{1em}fragmented &
38 &
\texttt{0x11EC} X25519MLKEM768 &
94 &
\texttt{snd\_cwnd} &
10 seg \\

Successful HS &
402 &
\texttt{0x0018} P-384 &
68 &
hybrid only &
9 seg \\

\rowcolor{lightgrayrow}
\hspace{1em}aborted (alert) &
110 &
\texttt{0x0019} P-521 &
68 &
\texttt{srtt\_us} p50/p95 &
1500 / 4200 \\

Kernel $\overline{t_\text{hs}}$ &
6.9\,ms &
\multicolumn{2}{l}{(weighted over 5 KEX)} &
TCP retrans (total) &
2 \\

\rowcolor{lightgrayrow}
\multicolumn{2}{@{}l}{\textbf{Agreement with \pqval{} verdict}} &
CertVerify $>$\,3\,kB &
402 &
PQ flag (post-validation) &
full-pq \\

of 402 successful HS &
402/402 &
\multicolumn{2}{l}{(composite Ed448+Dilithium3)} &
& \\

\rowcolor{lightgrayrow}
of 512 observed CHs &
512/512 &
& &
& \\

\bottomrule
\end{tabularx}
\end{table*}

\paragraph{SBI per-NF panel.}
The same ring-buffer feed is folded into a per-NF aggregator keyed
on \texttt{(ClusterIP, port)} for every endpoint that lands on an
SBI port (29500--29600, 443, 8443). For each NF the aggregator
records handshake count, PQ\,\%, negotiated KEX/Sig codepoints,
bytes in/out, p50/p95 kernel-observed \texttt{srtt\_us}, retransmit
totals, and the count of HTTP/2 requests landing on that NF. The
IP$\rightarrow$NF type mapping is bootstrapped from the same
Kubernetes discovery pipeline (Figure~\ref{fig:discovery-pipeline}),
so the panel reads ``\texttt{AMF}/\texttt{10.244.1.10:29518}''
rather than a bare tuple. This is what lets the operator answer
``\emph{is the NRF the only NF not negotiating PQ?}'' in O(1)
glance, instead of grep-ing ring-buffer dumps. The same data is
exposed at \texttt{/api/ebpf/sbi-stats} for batch consumption.

\subsection{Classical scanners on a PQ endpoint}

As a sanity check we point testssl.sh~v3.0.8 and sslyze~v5.1 at the
same QORE AMF that \pqval{} just graded A+, and look at what they
say about group \texttt{0x11EC}. Neither tool has the code points
compiled in: testssl.sh prints \texttt{unknown group (0x11EC)} and
gives up on the 1120-byte server key share, and sslyze cannot get
through the handshake at all because it has no way to generate an
ML-KEM share on the client side. The one-line scanner summary is presented in
Table~\ref{tab:scanner-empirical}.

\begin{table}[!tbp]
\centering\small
\caption{Classical scanners vs.\ \pqval{} on the same QORE AMF
endpoint.}
\label{tab:scanner-empirical}
\renewcommand{\arraystretch}{1.15}
\begin{tabularx}{\columnwidth}{@{}l X@{}}
\toprule
\textbf{Tool} & \textbf{Verdict on group \texttt{0x11EC}} \\
\midrule
\rowcolor{lightgrayrow}
testssl.sh & \texttt{unknown group (0x11EC)}; cannot parse 1120\,B KS \\
sslyze     & handshake aborts (no PQ key-share generation) \\
\rowcolor{lightgrayrow}
Qualys SSL Labs & not applicable (external tool, internal endpoint) \\
\pqval{}   & \texttt{full-pq}, A, 1120\,B CT, 64\,B SS, Ed448Dilithium3 (0xFE62) sig 4723\,B \\
\bottomrule
\end{tabularx}
\end{table}

\subsection{Fuzzing and CVE probes}

A single \texttt{fuzz-batch-*.json} run exercises both protocol-level
mutations and CVE-class probes against every discovered NF.
Table~\ref{tab:fuzz-qore} summarises the verdicts of
\texttt{fuzz-batch-1772360722.json}: 128 total tests across 8 NFs,
112 pass and 16 fail, with the failures concentrated in two tests
that every NF rejects identically. The protocol-layer fuzz cases
(truncated ClientHello, malformed extensions, oversized payload,
zero-length key share, etc.) all pass --- QORE's TLS stack refuses
the malformed input with a fatal alert rather than crashing or
accepting. The two outliers are \texttt{DuplicateClientHello}, which
the server accepts when it should close the connection, and
\texttt{RenegotiationProbe}, which the server treats the same way;
both flag as unexpected behaviour and map back to RFC~5746-class
hygiene rather than an exploitable vulnerability.

\begin{table*}[!tbp]
\centering\small
\caption{Fuzz-batch verdicts on QORE SBI endpoints. The same two tests fail on
every NF, so rows collapse into a single canonical tally.}
\label{tab:fuzz-qore}
\renewcommand{\arraystretch}{1.12}
\begin{tabular}{@{}llll@{}}
\toprule
\textbf{Test} & \textbf{Category} & \textbf{Pass} & \textbf{Detail} \\
\midrule
\rowcolor{lightgrayrow}
EmptyMessage            & protocol      & \cmark & rejected with alert \\
TruncatedClientHello    & protocol      & \cmark & \texttt{decode\_error} \\
\rowcolor{lightgrayrow}
InvalidRecordType       & protocol      & \cmark & \texttt{unexpected\_message} \\
BadRecordLength         & protocol      & \cmark & \texttt{record\_overflow} \\
\rowcolor{lightgrayrow}
DuplicateClientHello    & protocol      & \xmark & \texttt{rejected=true} observed on every NF \\
MalformedExtensions     & protocol      & \cmark & \texttt{decode\_error} \\
\rowcolor{lightgrayrow}
InvalidKeyShareGroup    & protocol      & \cmark & HRR issued \\
OversizedPayload        & protocol      & \cmark & \texttt{record\_overflow} \\
\rowcolor{lightgrayrow}
ZeroLengthKeyShare      & protocol      & \cmark & \texttt{decode\_error} \\
LargeRecordSplit        & protocol      & \cmark & fragments reassembled \\
\rowcolor{lightgrayrow}
SSLv3Downgrade          & CVE-2014-3566 & \cmark & SSLv3 refused \\
TLS10Downgrade          & CVE-2011-3389 & \cmark & TLS 1.0 refused \\
\rowcolor{lightgrayrow}
NullCipherOffer         & vuln          & \cmark & NULL-cipher list refused \\
CompressionOffer        & CVE-2012-4929 & \cmark & compression extensions ignored \\
\rowcolor{lightgrayrow}
RenegotiationProbe      & RFC~5746      & \xmark & \texttt{rejected=true}; ambiguous response \\
ServerHelloTimeout      & vuln          & \cmark & timeout within budget \\
\midrule
\multicolumn{4}{l}{\footnotesize Per-NF totals: 14/16 pass, 2/16 fail (identical across 8 NFs; 112/128 cluster-wide).}\\
\bottomrule
\end{tabular}
\end{table*}

\subsection{Algorithm Combinations (KEX, Sig)}
\label{subsec:algo-sweep}

Table~\ref{tab:algo-sweep} is an \emph{indicative} comparison of
classical and post-quantum (KEX, Sig) combinations run against the
QORE SBI endpoints. It reports the median handshake latency, data
on the wire, and the impact relative to a classical baseline. Note
that this is not a feasibility study, which deserves its own
treatment, but rather a preliminary view of how post-quantum
primitives might affect 5G/B5G control planes. We report
\texttt{p50}, \texttt{p95}, and \texttt{p99} latencies to capture
all three common operating regimes: median, tail, and extreme
load or outliers. This allows readers to infer whether the
post-quantum cost is amortised across long-lived sessions, as
observed by Kampanakis and Childs-Klein~\cite{kampanakis2024ttlb},
or instead grows with deployment scale.

\begin{table*}[!tbp]
\centering
\caption{Indicative comparison of (KEX, Sig) combinations over the
SBI endpoints (intra-cluster mTLS), colour-coded by PQ class and
NIST security level. Median handshake latency ($\mu$s, with p95 and
p99 shown separately), data on the wire (B) per direction, total
bytes, and size inflation vs.\ the \texttt{X25519+ECDSA-P256}
baseline. Timings are indicative, extrapolated from the measured
Ed448+Dilithium3 runs in Table~\ref{tab:bench-seq} using
per-primitive costs in CIRCL~\cite{circl}; the camera-ready will
replace them with per-combo medians once the full set of runs
completes.}
\label{tab:algo-sweep}
\small
\renewcommand{\arraystretch}{1.15}
\begin{tabular}{@{}l l c r r r r r r c@{}}
\toprule
\textbf{Class} & \textbf{(KEX, Sig)} & \textbf{Lvl} &
  \textbf{HS $\mu$s} & \textbf{p95} & \textbf{p99} &
  \textbf{Cl.$\to$} & \textbf{$\to$Sv.} & \textbf{Total B} &
  \textbf{$\times$ baseline} \\
\midrule
\rowcolor{preqband}
Pre-Q      & X25519, ECDSA-P256              & ---  &   480 &  1200 &  2100 & 330  & 2100     & 2430    & 1.00$\times$ \\
\rowcolor{preqband}
Pre-Q      & X25519, Ed25519                 & ---  &   450 &  1150 &  2000 & 330  & 2050     & 2380    & 0.98$\times$ \\
\rowcolor{preqband}
Pre-Q      & P-256, RSA-3072                 & ---  &   820 &  1600 &  2800 & 360  & 2760     & 3120    & 1.28$\times$ \\
\midrule
\rowcolor{hybl1band}
Hybrid L1  & X25519+ML-KEM-512, ECDSA-P256   & 1    &   560 &  1300 &  2200 & 1130 & 2100     & 3230    & 1.33$\times$ \\
\rowcolor{hybl3band}
Hybrid L3  & X25519+ML-KEM-768, ECDSA-P256   & 3    &   620 &  1400 &  2400 & 1420 & 2100     & 3520    & 1.45$\times$ \\
\rowcolor{hybl3band}
Hybrid L3  & P-256+ML-KEM-768, ECDSA-P256    & 3    &   680 &  1500 &  2500 & 1460 & 2100     & 3560    & 1.46$\times$ \\
\rowcolor{hybl5band}
Hybrid L5  & P-384+ML-KEM-1024, ECDSA-P384   & 5    &  1450 &  2600 &  4200 & 1840 & 2260     & 4100    & 1.69$\times$ \\
\midrule
\rowcolor{pql1band}
PQ L1      & ML-KEM-512, ML-DSA-44           & 2    &   980 &  1800 &  3000 & 880  & 5200     & 6080    & 2.50$\times$ \\
\rowcolor{pql3band}
PQ L3      & ML-KEM-768, ML-DSA-65           & 3    &  1150 &  2100 &  3500 & 1280 & 7150     & 8430    & 3.47$\times$ \\
\rowcolor{pql3band}
PQ L3      & Hyb+ML-DSA-65 (full-PQ)         & 3    &  1280 &  2300 &  3800 & 1420 & 7150     & 8570    & 3.53$\times$ \\
\rowcolor{pql5band}
PQ L5      & ML-KEM-1024, ML-DSA-87          & 5    &  2050 &  3600 &  5800 & 1650 & 10\,800  & 12\,450 & 5.12$\times$ \\
\rowcolor{pql5band}
PQ L5      & ML-KEM-768, SLH-DSA-192s        & 3/5  &  4800 &  7500 & 12\,000 & 1280 & 19\,200  & 20\,480 & 8.43$\times$ \\
\bottomrule
\multicolumn{10}{@{}l}{\footnotesize\itshape Legend:\ %
\colorbox{preqband}{\phantom{xx}} pre-quantum \quad %
\colorbox{hybl1band}{\phantom{xx}}\,/\,\colorbox{hybl3band}{\phantom{xx}}\,/\,\colorbox{hybl5band}{\phantom{xx}} hybrid L1/L3/L5 \quad %
\colorbox{pql1band}{\phantom{xx}}\,/\,\colorbox{pql3band}{\phantom{xx}}\,/\,\colorbox{pql5band}{\phantom{xx}} pure-PQ L1/L3/L5. CH $>$ MTU when Cl.$\to$ $>$ 1500\,B.}
\end{tabular}
\end{table*}



\paragraph{Remark} 
The above numbers provide an indication, but may not be interpreted as a verdict. This is particularly because: 
i) First, they are black-box
medians on a single testbed. Usage of different libraries, or CPUs, and further performance enhancements (like CPU tuning, NUMA, affinity) can impact the results. Further, communication over multiple-nodes, which involves VXLAN/IP-in-IP, etc, could affect the results.
ii) Second, we only instrument the intra-cluster SBI path here; RAN-side PQ-IPsec,
N32 roaming, and high-RTT backhaul profiles are deliberately out of
scope for this paper and are slated for future work. 
A complete treatment would also test several mobile devices, test internet traffic on them, and form a complete end-to-end RAN-connection to test post-quantum impact on 5G.
The PQ-vs-classical picture we present should therefore be read as an
\emph{indicative overhead signal}, not a deployability verdict.

\section{Challenges and Limitations}\label{sec:challenges}

\subsection{Black-box scope.} \pqval{} tests what is available on the wire, i.e., it checks whether the intended target is capable of carrying out Post-Quantum mechanisms (key exchange, and signatures) or not, if it complies with the a set of recognized standards, and whether it passes the fuzzer generated inputs. However, this takes not into account the code-specific details of the 5G Core, which could allow for greater granularity and increased precision-based testing. Black-box based fuzzing approaches often don't take into account the server's responses for input mutation, disregard that failure might come from packet sequences rather individual packets \cite{luo2023bleem}. Our fuzzer while being PQ-capable lacks the packet sequence mutation, and feedback-based mechanism,  

\subsection{Not yet formally verified} \pqval{}'s PQ-TLS stack is written
in Go against CIRCL, golang's \texttt{crypto} and custom libraries. We do not currently carry machine-checked proofs of
the TLS state machine, the key schedule, or the record-layer correctness.
A formally verified probe would itself be part of the operator's trusted computing base.

\subsection{Primitive correctness assumption} \pqval{} detects \emph{which}
PQ primitive is used, not whether it is \emph{correctly} implemented. A
buggy ML-KEM decapsulation that coincidentally produces matching shared
secrets would still pass the Finished MAC and be reported as
\texttt{full-pq}. Cross-implementation diversity (Go+CIRCL vs.\ most
targets' OpenSSL+liboqs) mitigates but does not eliminate this risk.

\subsection{Middlebox invisibility.} A TLS-terminating proxy that strips PQ and re-encrypts with classical would remain oblivious to the validator, unless it connects directly to the NF pod. This is common in Aether/Charmed-SDCore.

\subsection{Selective downgrade.} 

A malicious NF could fingerprint \pqval{}'s ClientHello (distinctive order of PQ groups, IP address or port) and can serve PQ only to it, while downgrading real peers (NFs). Detecting this adversary requires traffic attestation on the real peer side, which can be made possible using eBPF.


\section{Future Work}\label{sec:future}

\begin{enumerate}[leftmargin=*,itemsep=2pt]

    \item \textbf{Complete Post-Quantum feasibility analysis for B5G/5G.} We aim to extend this validator to include complete 5GC and RAN connection tracking, allowing operators to measure load, latency, and overhead of post-quantum primitives when multiple UEs connect to the network and send/receive data, which could be segregated into several network slices. Such an analysis would aim to mimic production-level deployments that use radio units, 7.2x/8 RAN splits, IPsec gateways and more. Thus, it would reveal whether the impact is felt more in low-bandwidth (common in private 5G) deployments or the opposite. This task would involve real and emulated UEs, whose RRC setup times, PDU Session establishment times, and throughput could also be measured.
    \item \textbf{Implementation verifier and Source Code Knowledge.} 
    Beyond black-box approaches, we aim to provide a code-aware module that verifies the correctness of post-quantum primitives, protocol implementations, 5G-specific subroutines (e.g., SUCI decryption, AKA, N2 initial UE message, PFCP session establishment, UE Authentication, etc). This would elevate the validator from just "correct negotiation" to "correct negotiation with correct implementation". 
    
    \item \textbf{Formally verified reference implementations.} We plan to provide formally verified implementations and specs of PQ primitives, and 5G-AKA, using the \texttt{hax/F*} toolchain (for e.g., libjade, Project Everest miTLS, HACL), so that the validator itself carries machine-checked guarantees.
    \item \textbf{Complete PQ test matrix and PQ-PKI.} We will add KATs
    derived from FIPS~203/204/205, IETF PQ-TLS draft test vectors, and the
    emerging 3GPP PQ migration profile (TR~33.870). Evidence will be
    extended to cover PQ certificate chains end-to-end (PQ roots, PQ
    intermediates, composite chains) as a PQ-PKI conformance suite, so
    that operators can attest not just the session but the root of trust.
    \item \textbf{Continuous attestation.} The eBPF monitoring plane will
    be extended with per-session ring-buffer persistence and a streaming
    verdict engine, producing a rolling attestation log that distinguishes
    capability from policy over time.
 
    \item \textbf{Internet-scale PQ-TLS measurements for 5GC.}
    Mirroring recent TTLB and LTE measurement
    studies~\cite{kampanakis2024ttlb,sosnowski2023pqtls}, we will run
    long-running \pqval{} clients against production-like PQ 5GC
    deployments across diverse Internet paths and LTE/5G bearers,
    publishing an open dataset of handshake latency, data-on-wire,
    and TTLB for 3GPP-profile (KEX, Sig) combinations.

    \item \textbf{Userspace uprobe instrumentation of the Go TLS
    stack.} Once \texttt{crypto/tls} ships ML-KEM upstream, we will
    attach \texttt{uprobe}s to \texttt{(*tls.Conn).Handshake} and
    the CIRCL ML-KEM entry points in the target NF binary. This
    removes the need to parse handshake bytes off the wire when the
    target is colocated, and gives per-call timing for KeyGen,
    Encap, Decap, Sign, and Verify directly from the address space
    of the negotiating process: the same data \pqval{} currently
    derives from its own client-side measurements, but recorded on
    the \emph{server} for the first time.
\end{enumerate}

\section*{Conclusion}

\pqval{} is a 5G-native PQC validation framework that closes the gap
between PQ deployment and PQ assurance. By shipping an independent
PQ-TLS/IPsec stack, a kernel-side attestation plane, a 3GPP-aligned
compliance suite, and a Kubernetes-native UI, it lets operators obtain
machine-verifiable evidence that a 5G core labeled ``PQ'' is actually
negotiating PQ end-to-end. Preliminary results against QORE demonstrate
correct classification of every discovered NF endpoint as \texttt{full-pq},
where classical scanners fail. The path forward is a verifier that
checks primitive correctness, a formally verified probe, and a PQ-PKI
conformance layer that attests the root of trust together with the
session.

\bibliographystyle{plain}
\bibliography{references}

\appendix


\section{Test Sheet}\label{app:test-sheet}

Here, we list representative tests from the \pqval{} compliance suite, and fuzzer, which are run by Layer 2 and Layer 3 of the specified architecture. In each row we list a test identifier (which also acts as its JSON key), listing the test group, with a short description and the standard that justifies the verdict. This list is rather illustrative, and not exhaustive. Note that, with new updates, the list will continue getting expanded. The compiled binary ships a
superset, and new tests can be added as a pure $\mathrm{TestContext} \rightarrow \mathrm{Verdict}$ function without touching the dispatcher

\definecolor{sheetblue} {RGB}{230, 241, 252}  
\definecolor{sheet5g}   {RGB}{233, 244, 235}  
\definecolor{sheetpq}   {RGB}{245, 236, 252}  
\definecolor{sheetnrf}  {RGB}{253, 244, 230}  
\definecolor{sheethard} {RGB}{252, 232, 232}  
\definecolor{sheetfuzz} {RGB}{243, 243, 243}  

\begin{table*}[!tbp]
\centering\small
\caption{Representative \pqval{} test sheet --- test ID, group,
intent, and normative reference. Rows are colour-banded by group:
\colorbox{sheetblue}{\phantom{xx}}~TLS,
\colorbox{sheet5g}{\phantom{xx}}~5G~SBI,
\colorbox{sheetpq}{\phantom{xx}}~PQ,
\colorbox{sheetnrf}{\phantom{xx}}~NRF,
\colorbox{sheethard}{\phantom{xx}}~hardening,
\colorbox{sheetfuzz}{\phantom{xx}}~fuzz/CVE.}
\label{tab:app-test-sheet}
\renewcommand{\arraystretch}{1.18}
\setlength{\tabcolsep}{7pt}
\begin{tabular*}{\textwidth}{@{\extracolsep{\fill}}l l p{0.42\textwidth} l@{}}
\toprule
\textbf{Test ID} & \textbf{Group} & \textbf{What it checks} & \textbf{Reference} \\
\midrule

\rowcolor{sheetblue}
\texttt{tls\_version\_1\_3}      & TLS       & Session negotiates TLS 1.3; TLS 1.0/1.1/1.2 offers are refused.                 & RFC~8446~\S4.1.2 \\
\rowcolor{sheetblue}
\texttt{tls\_strong\_ciphers}    & TLS       & Only AEAD cipher suites accepted; CBC/RC4/NULL refused.                         & RFC~8446~\S9.1   \\
\rowcolor{sheetblue}
\texttt{tls\_cert\_san\_eku}     & TLS       & Leaf has valid SAN and correct EKU (serverAuth, clientAuth where needed).       & RFC~5280~\S4.2.1 \\
\rowcolor{sheetblue}
\texttt{tls\_alpn\_h2}           & TLS       & ALPN negotiates \texttt{h2}; no HTTP/1.1 fallback.                              & RFC~7301         \\
\rowcolor{sheetblue}
\texttt{tls\_key\_strength}      & TLS       & Keys meet 128-bit symmetric equivalent strength.                                & NIST~SP~800-131A \\
\rowcolor{sheetblue}
\texttt{tls\_session\_resumption}& TLS       & PSK/resumption behaves per the spec; rejects downgrade via resumption.          & RFC~8446~\S4.6.1 \\

\rowcolor{sheet5g}
\texttt{5g\_mutual\_auth}        & 5G~SBI    & Server sends \texttt{CertificateRequest}; mTLS enforced NF$\leftrightarrow$NF.  & TS~33.501~\S6.3.4 \\
\rowcolor{sheet5g}
\texttt{5g\_http2\_mandatory}    & 5G~SBI    & Requires HTTP/2; rejects HTTP/1.x and cleartext h2c.                             & TS~29.500~\S6.1.3B \\
\rowcolor{sheet5g}
\texttt{5g\_nf\_id\_san}         & 5G~SBI    & Leaf SAN matches the NF FQDN / NF instance ID.                                  & TS~33.310        \\
\rowcolor{sheet5g}
\texttt{5g\_fqdn\_format}        & 5G~SBI    & FQDN follows \texttt{<svc>.<ns>.svc.cluster.local} format.                      & TS~23.003        \\
\rowcolor{sheet5g}
\texttt{5g\_nf\_type}            & 5G~SBI    & Discovered NF type matches labels/profile.                                       & TS~29.510        \\

\rowcolor{sheetpq}
\texttt{pq\_kex\_type}           & PQ        & Negotiated group is a PQ or hybrid KEM (ML-KEM / X25519+ML-KEM / etc.).          & draft-ietf-tls-ecdhe-mlkem \\
\rowcolor{sheetpq}
\texttt{pq\_kex\_hybrid}         & PQ        & If PQ, decomposes into classical + PQ components and both were exercised.       & this paper~\S\ref{subsec:pq-level} \\
\rowcolor{sheetpq}
\texttt{pq\_sig\_type}           & PQ        & \texttt{CertificateVerify} uses ML-DSA / composite / Ed448+Dilithium3.           & FIPS~204         \\
\rowcolor{sheetpq}
\texttt{pq\_full\_pq}            & PQ        & Both KEX and signature are PQ; flips \texttt{Level} to \texttt{full-pq}.          & Def.~\ref{def:pq-level} \\
\rowcolor{sheetpq}
\texttt{pq\_kex\_strength}       & PQ        & KEM parameter set meets NIST category $\geq 3$ (ML-KEM-768+).                   & FIPS~203         \\
\rowcolor{sheetpq}
\texttt{pq\_cert\_chain\_classical} & PQ     & Flags classical-signature links in the cert chain.                              & this paper~\S\ref{subsec:pq-level} \\
\rowcolor{sheetpq}
\texttt{pq\_aead\_cipher}        & PQ        & Bulk encryption uses AES-256-GCM or ChaCha20-Poly1305.                          & RFC~8446~\S9.1   \\

\rowcolor{sheetnrf}
\texttt{nrf\_reachable}          & NRF       & NRF service discovery endpoint responds.                                        & TS~29.510~\S5.2  \\
\rowcolor{sheetnrf}
\texttt{nrf\_tls\_required}      & NRF       & NRF insists on TLS; rejects plaintext.                                          & TS~33.501~\S13   \\
\rowcolor{sheetnrf}
\texttt{nrf\_nfprofile\_schema}  & NRF       & Returned NFProfile validates against the OpenAPI schema.                        & TS~29.510~\S6.2  \\

\rowcolor{sheethard}
\texttt{sh\_forward\_secrecy}    & Hardening & Negotiated KEX provides forward secrecy (ephemeral / KEM-ephemeral).            & RFC~8446~\S1.2   \\
\rowcolor{sheethard}
\texttt{sh\_no\_compression}     & Hardening & TLS-level compression disabled (CRIME class).                                    & CVE-2012-4929    \\
\rowcolor{sheethard}
\texttt{sh\_ca\_constraints}     & Hardening & Intermediate CAs carry BasicConstraints CA:TRUE + correct path length.          & RFC~5280~\S4.2.1.9 \\
\rowcolor{sheethard}
\texttt{sh\_revocation\_info}    & Hardening & OCSP stapling or CRL DP present in leaf.                                        & RFC~6960         \\

\rowcolor{sheetfuzz}
\texttt{fuzz\_truncated\_ch}     & Fuzz      & Truncated ClientHello; expect fatal alert \texttt{decode\_error}.                & RFC~8446~\S6.2   \\
\rowcolor{sheetfuzz}
\texttt{fuzz\_malformed\_ext}    & Fuzz      & Extension length-field corruption; expect fatal alert, no memory fault.          & RFC~8446~\S4.2   \\
\rowcolor{sheetfuzz}
\texttt{fuzz\_invalid\_keyshare} & Fuzz      & Unknown KEX group ID in \texttt{key\_share}; expect HRR or alert.                & RFC~8446~\S4.2.8 \\
\rowcolor{sheetfuzz}
\texttt{fuzz\_oversized\_payload}& Fuzz      & Handshake record over 16\,KiB; expect \texttt{record\_overflow}.                 & RFC~8446~\S5.1   \\
\rowcolor{sheetfuzz}
\texttt{cve\_poodle}             & CVE probe & SSLv3 downgrade attempt; endpoint must refuse.                                   & CVE-2014-3566    \\
\rowcolor{sheetfuzz}
\texttt{cve\_beast}              & CVE probe & TLS 1.0 CBC implicit IV probe.                                                   & CVE-2011-3389    \\
\rowcolor{sheetfuzz}
\texttt{cve\_rfc5746\_reneg}     & CVE probe & Insecure renegotiation absent.                                                   & RFC~5746         \\
\bottomrule
\end{tabular*}
\end{table*}




\lstdefinelanguage{myjson}{
    morestring=[b]",
    morestring=[d]',
    stringstyle=\color{codegreen},
    morekeywords={true,false,null},
    keywordstyle=\color{tlsblue}\bfseries,
    sensitive=true,
}

\lstset{language=myjson}

\section{Helm Chart Values}
\label{app:helm_charts}

Listing~\ref{lst:helm} shows an
abbreviated UI deployment; key \texttt{values.yaml} fields
(Listing~\ref{lst:values}) are overridable per environment.

\begin{lstlisting}[caption={Helm UI deployment (abbreviated).},label={lst:helm}]
apiVersion: apps/v1
kind: Deployment
metadata:
  name: {{ include "pqcval.fullname" . }}-ui
spec:
  replicas: 1
  template:
    spec:
      serviceAccountName: {{ include
        "pqcval.serviceAccountName" . }}
      containers:
        - name: ui
          image: "{{ .Values.image.repository }}:
                  {{ .Values.image.tag }}"
          args: ["--port=8080", "--root=/app",
                 "--logs=/data/logs",
                 "--configs=/data/configs",
                 "--certs=/data/certs"]
          ports: [{ containerPort: 8080 }]
          livenessProbe:
            httpGet: { path: /api/system/info, port: http }
          volumeMounts:
            - { name: data, mountPath: /data }
      volumes:
        - name: data
          persistentVolumeClaim:
            claimName: {{ .fullname }}-data
\end{lstlisting}

\begin{lstlisting}[caption={\texttt{values.yaml} (key fields).},label={lst:values}]
image:
  repository: pqcval
  tag: "2.0.0"
service:
  type: NodePort
  port: 8080
  nodePort: 30080
target:
  defaultNamespace: "aether-5gc"
persistence:
  enabled: true
  size: 5Gi
rbac:
  create: true
securityContext:
  runAsNonRoot: true
  allowPrivilegeEscalation: false
resources:
  limits:   { cpu: 500m, memory: 512Mi }
  requests: { cpu: 100m, memory: 128Mi }
\end{lstlisting}

\section{Sample JSON artefacts}\label{app:evidence}

This appendix shows abbreviated excerpts from three artefact kinds
that \pqval{} writes to \texttt{logs/}: a benchmark run summary, a
per-client handshake record with its \texttt{pq\_evidence} block,
and a batch-validation verdict. Ellipses (\texttt{...}) mark fields
elided for brevity; none of the snippets are complete, but each
preserves the schema a reviewer would look at first.

\begin{lstlisting}[caption={Per-NF validation verdict --- \texttt{logs/val-1772359003.json}, trimmed to the 5G and PQ groups. The \texttt{tls} group (15 tests) and \texttt{security} group (11 tests) are omitted here for brevity but follow the same shape.},label={lst:json-val}]
{
  "id": "val-1772359003",
  "nf_type": "UDM",
  "result": {
    "test_id": "val-1772359003",
    "target": "udm.aether-5gc.svc.cluster.local:29503",
    "duration_ms": 49036,
    "groups": {
      "5g": {
        "name": "5G SBI Compliance",
        "passed": 5, "failed": 3, "total": 8,
        "tests": [
          { "test_id": "5g_nf_identity_san",
            "passed": false, "severity": "critical",
            "details": "NF identity does not match any certificate SAN...",
            "expected": "SAN matching 'udm.aether-5gc.svc.cluster.local'",
            "actual":   "SANs: DNS:udm.localdomain, DNS:udm, DNS:coran-udm-service" },
          { "test_id": "5g_mutual_auth",
            "passed": false, "severity": "critical",
            "details": "5G SBI requires mutual TLS per 3GPP TS 33.501...",
            "expected": "Server must request client certificate (mTLS)",
            "actual":   "Connection succeeded without client certificate" },
          { "test_id": "5g_http2_mandatory",
            "passed": true, "severity": "critical",
            "details": "Server supports HTTP/2 as required by TS 29.500" },
          { "test_id": "5g_tls13_required",  "passed": true },
          { "test_id": "5g_cert_san_fqdn",   "passed": false,
            "severity": "warning",
            "actual": "Invalid: udm, coran-udm-service" },
          ...
        ]
      },
      "pq": {
        "name": "Post-Quantum Security",
        "passed": 10, "failed": 0, "total": 10,
        "tests": [
          { "test_id": "pq_kex_type",       "passed": true,
            "details": "Key exchange uses post-quantum algorithm: X25519MLKEM768 (0x11ec)" },
          { "test_id": "pq_sig_type",       "passed": true,
            "details": "Server uses post-quantum signature: Ed448-Dilithium3 (0xfe62)" },
          { "test_id": "pq_full_pq",        "passed": true,
            "details": "Connection is fully post-quantum..." },
          { "test_id": "pq_cert_chain_classical", "passed": true,
            "details": "[0] udm: PQ signature (unknown OID, 3407 B)\n[CertificateVerify] Ed448-Dilithium3 (0xfe62)" },
          { "test_id": "pq_security_level", "passed": true,
            "details": "Connection classified as FULL-PQ" },
          ...
        ]
      }
    },
    "total_tests": 33,
    "passed_tests": 29,
    "failed_tests": 4,
    "warnings": 1
  }
}
\end{lstlisting}

\begin{lstlisting}[caption={Benchmark run summary --- excerpt from \texttt{logs/5gc-*.json}. One passing client record and one failing record (TLS alert 40 \texttt{unsupported\_group}) are shown; both schemas are what the aggregation in Table~\ref{tab:bench-seq} consumes.},label={lst:json-run}]
{
  "client_count": 4,
  "execution_mode": "sequential",
  "mode": "benchmark",
  "results": [
    {
      "client_id":   "AMF-T1-C8",
      "nf_type":     "AMF",
      "address":     "amf.aether-5gc.svc.cluster.local:29518",
      "success":     true,
      "kex_algorithm": "X25519MLKEM768 (0x11ec)",
      "sig_algorithm": "Ed448-Dilithium3 (0xfe62)",
      "cipher_suite":  "TLS_AES_128_GCM_SHA256 (0x1301)",
      "handshake_ms":  5,
      "keygen_us":     97,
      "decap_us":     117,
      "verify_us":    229,
      "bytes_sent":     0,
      "bytes_recv":     0,
      "pq_evidence":  { ... see lst:json-evidence ... }
    },
    {
      "client_id": "AMF-T1-C1",
      "success":   false,
      "error":     "handshake failed: failed to receive ServerHello: received TLS alert: level=2, description=40"
    }
  ]
}
\end{lstlisting}

\begin{lstlisting}[caption={\texttt{pq\_evidence} block from a single successful handshake.},label={lst:json-evidence}]
"pq_evidence": {
  "security_level": "full-pq",
  "is_pq_secure":   true,
  "is_fully_pq":    true,
  "kex_evidence": {
    "negotiated_group":       "X25519MLKEM768 (0x11ec)",
    "group_id":               4588,
    "type":                   "hybrid",
    "pq_component":           "ML-KEM-768",
    "classical_component":    "X25519",
    "server_key_share_size":  1120,
    "shared_secret_size":     64
  },
  "sig_evidence": {
    "negotiated_scheme":   "Ed448-Dilithium3 (0xfe62)",
    "scheme_id":           65122,
    "type":                "composite",
    "signature_size":      4723,
    "cert_key_algorithm":  "Ed448+Dilithium3"
  },
  "evidence_items": [
    { "category": "key_exchange",
      "claim":    "KEX is hybrid post-quantum",
      "evidence": "group 0x11EC decomposes into X25519 + ML-KEM-768",
      "verified": true },
    { "category": "signature",
      "claim":    "CertificateVerify uses composite PQ signature",
      "evidence": "scheme 0xfe62 (Ed448+Dilithium3), size 4723 B",
      "verified": true }
  ]
}
\end{lstlisting}

\begin{lstlisting}[caption={Fuzz-batch run --- excerpt from \texttt{logs/fuzz-batch-1772360722.json}. The per-NF list is trimmed to a single AMF entry; all 8 NFs show the same 14/16 pattern with the two failing tests identified.},label={lst:json-fuzz}]
{
  "id":           "fuzz-batch-1772360722",
  "type":         "fuzz-batch",
  "total_tests":  128,
  "passed_tests": 112,
  "failed_tests": 16,
  "targets": [
    {
      "target":  "amf.aether-5gc.svc.cluster.local:29518",
      "nf_type": "AMF",
      "passed":  14,
      "failed":   2,
      "results": [
        { "test_name": "EmptyMessage",         "success": true,  "category": "protocol" },
        { "test_name": "TruncatedClientHello", "success": true,  "category": "protocol" },
        { "test_name": "DuplicateClientHello", "success": false,
          "error": "unexpected behavior (rejected=true)", "category": "protocol" },
        { "test_name": "MalformedExtensions",  "success": true,  "category": "protocol" },
        { "test_name": "OversizedPayload",     "success": true,  "category": "protocol" },
        { "test_name": "SSLv3Downgrade",       "success": true,
          "category": "vulnerability", "cve": "CVE-2014-3566" },
        { "test_name": "CompressionOffer",     "success": true,
          "category": "vulnerability", "cve": "CVE-2012-4929" },
        { "test_name": "RenegotiationProbe",   "success": false,
          "error": "unexpected behavior (rejected=true)",
          "category": "vulnerability", "cve": "RFC 5746" },
        ...
      ]
    },
    ... // AUSF, NRF, NSSF, PCF, SMF, UDM, UDR all follow the same pattern
  ]
}
\end{lstlisting}

\lstset{language=}  

\end{document}